%% file: main.tex
\documentclass[11pt]{book}
\usepackage[table]{xcolor}
\usepackage{booktabs}
\usepackage{threeparttablex}
\usepackage{longtable}
\usepackage{Wiley-AuthoringTemplate}
\usepackage{chapterbib}
\usepackage[sectionbib,authoryear]{natbib}
\usepackage{latexsym}
\usepackage{graphicx}
\usepackage{amsfonts,amssymb,amsmath}
\usepackage{varwidth}
\usepackage{float}
\usepackage[switch]{lineno}

\usepackage{algorithm,algorithmic}  

\usepackage{subcaption}
\usepackage{comment}
\usepackage{amsthm}
\usepackage{overpic}
\usepackage{steinmetz}
\usepackage{array}
\usepackage{url}

\usepackage[T1]{fontenc}
\usepackage{cite}
\usepackage{caption}
\usepackage{hyperref}

\newcommand{\argmax}[1]{{\underset{{#1}}{\mathrm{arg\,max}}}}

\newcommand{\vect}[1]{\mathbf{#1}}
\newcommand{\bl}[1]{\boldsymbol{#1}}

\def\diag{\mathrm{diag}}

\def\F{\mathrm{F}}

\def\FA{\mathrm{FA}}
\def\T{\mathrm{T}}
\def\H{\mathrm{H}}

\def\CN{\mathcal{N}_{\mathbb{C}}} 

\makeindex
\begin{document}

\mainmatter
\include{Chapter}

\backmatter

\end{document}

%% file: Chapter.tex
\chapter{Localization in Massive MIMO Networks: From Near-Field to Far-Field   }

\author[1]{Parisa Ramezani}
\author[2]{\"Ozlem Tu\u{g}fe Demir}
\author[3]{Emil Bj\"ornson}

\address[1]{\orgdiv{Department of Computer Science}, 
\orgname{KTH Royal Institute of Technology},
    \city{Stockholm},
    \country{Sweden}}%

\address[2]{\orgdiv{Department of Electrical and Electronics Engineering}, 
\orgname{TOBB University of Economics and Technology}, 
     \city{Ankara}, \country{T\"urkiye}}%

\address*{Corresponding Author: \email{emilbjo@kth.se}}

\maketitle

\begin{abstract}{Abstract}
Source localization is the process of estimating the location of signal sources based on the signals received at different antennas of an antenna array. It has diverse applications, ranging from radar systems and underwater acoustics to wireless communication networks. Subspace-based approaches are among the most effective techniques for source localization due to their high accuracy, with Multiple SIgnal Classification (MUSIC) and Estimation of Signal Parameters by Rotational Invariance Techniques (ESPRIT) being two prominent methods in this category. These techniques leverage the fact that the space spanned by the eigenvectors of the covariance matrix of the received signals can be divided into signal and noise subspaces, which are mutually orthogonal. Originally designed for far-field source localization, these methods have undergone several modifications to accommodate near-field scenarios as well. This chapter aims to present the foundations of MUSIC and ESPRIT algorithms and introduce some of their variations for both far-field and near-field localization by a single array of antennas. We further provide numerical examples to demonstrate the performance of the presented methods. 
\end{abstract}
\keywords{Array signal processing, near-field localization, MUSIC, ESPRIT.}

\section{Introduction}

Antenna arrays with massive antenna numbers are commonplace at the base stations in 5G networks. These arrays are deployed to enable Massive MIMO (multiple-input multiple-output) communications between the multi-antenna base station and a large number of spatially multiplexed user devices \citep{bjornson2017massive}. However, these arrays might be used for additional services in the future \citep{chen20236g}, and this chapter focuses on localization services.

Source localization is an important problem in the field of array signal processing and has a wide range of applications in sonar, radar, seismic exploration, and wireless communications \citep{Wax1983,Ziskind1988,10144727,pesavento2023three,9132699,9960789}. As such, it has drawn considerable attention over the past decades and various algorithms have been developed to address this problem \citep{liang2009passive,6451125,8736783,9132699,9960789}. Subspace-based approaches are among the most popular and widely used classical techniques for source localization when using a single antenna array \citep{pesavento2023three}. These approaches utilize the properties of the signal and noise subspaces of the received signals to estimate the locations of the sources. In particular, they are built on the fact that the signal subspace is generally of lower dimension than the total measurement space which helps in separating the signal from the noise. Subspace-based methods often provide high spatial resolution, meaning that they can accurately estimate the location of closely spaced sources. However, their performance is dependent on various factors such as the array geometry, signal-to-noise ratio (SNR), number of antennas (sensors), etc. Two classical subspace-based localization methods are  MUltiple SIgnal Classification (MUSIC) \citep{Schmidt1986,stoica2005spectral} and Estimation of Signal Parameters by Rotational Invariance Techniques (ESPRIT) \citep{Roy1986,stoica2005spectral}, which respectively exploit the noise and signal subspace of the received signals for localizing the sources. 

When considering large antenna arrays, the sources can be classified into two categories based on their distance from the array: far-field sources and near-field sources. Far-field sources are characterized by planar wavefronts, where the curvature of the wave arriving at the antenna array is negligible and the localization only involves estimating the direction of arrival (DoA) of the source. When the source is located in the radiative near-field region of the antenna array, the spherical wavefront of the received signal can no longer be neglected and the locations of the sources are specified by both the DoA and range \citep{Rockah1987}. 

\subsection{Key Contributions}
This chapter provides an overview of source localization considering both far-field and near-field scenarios. We first present the popular MUSIC and ESPRIT algorithms for cases where the sources are situated in the far-field of the antenna array. We then describe the generalized ESPRIT approach, which relaxes the strict shift-invariance requirement of the conventional ESPRIT algorithm and is applicable to a general class of subarray geometries. Afterward, we describe the application of MUSIC and ESPRIT algorithms to near-field localization, in which both the DoA and range must be estimated for localizing the sources. The two-dimensional (2D) MUSIC is first put forth, which simultaneously estimates the DoAs and ranges via a 2D grid search over all possible angles and ranges. This grid search can lead to high computational costs, especially when the grid resolution is fine. To alleviate this issue, some algorithms have been developed to enable the separate estimation of DoAs and ranges. We present two methods based on MUSIC and ESPRIT algorithms that exploit the symmetric geometry of the antenna array to decouple the DoA and range estimation problems. These methods first estimate the DoAs of all sources via a one-dimensional (1D) search over possible angles and then perform multiple 1D searches to obtain the range associated with each estimated DoA. In this way, the location of sources can be estimated through multiple 1D searches, thereby reducing the complexity of the 2D search required by the conventional MUSIC algorithm.

\subsection{Chapter Organization}
In Section~\ref{sec:far-field}, we introduce MUSIC and ESPRIT algorithms for far-field localization and also present the generalized ESPRIT method that extends the standard ESPRIT algorithm to a more general class of array geometries. Section~\ref{sec:near-field} discusses near-field localization and clarifies how the standard 1D MUSIC can be extended to 2D MUSIC for locating near-field sources. This section also introduces symmetry-based localization where the symmetric structure of the array is exploited to decouple DoA and range estimation problems in the near-field. Finally, Section~\ref{sec:conclusion} provides concluding remarks.

\subsection{Notations}
Scalars are denoted by italic letters, vectors and matrices are denoted by bold-face lower-case and upper-case letters, respectively. $\mathbb{C}$ represents the set of complex numbers and $\CN (0,\sigma^2)$ indicates a circularly symmetric complex Gaussian distribution with variance $\sigma^2$. $(\cdot)^*$ indicates conjugation, and $(\cdot)^\T$, $(\cdot)^\H$, and $(\cdot)^{-1}$ represent transpose, conjugate transpose, and inverse of a matrix, respectively. $\diag(\vect{x})$ represents a diagonal matrix having $\vect{x}$ on its main diagonal, and $\det(\cdot)$ denotes the determinant of a matrix.

\section{Far-Field DoA Estimation}
\label{sec:far-field}
In this section, we review the conventional MUSIC and ESPRIT algorithms for locating far-field sources and numerically evaluate their performance. We further introduce the generalized ESPRIT algorithm that extends the concept of standard ESPRIT by allowing for arbitrary displacements between the corresponding antennas of the subarrays.

\subsection{System Model}
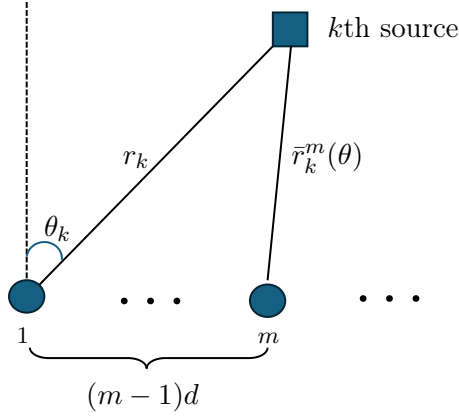
\begin{figure}[t!]
  \centering
   \begin{overpic}[scale = 0.5]{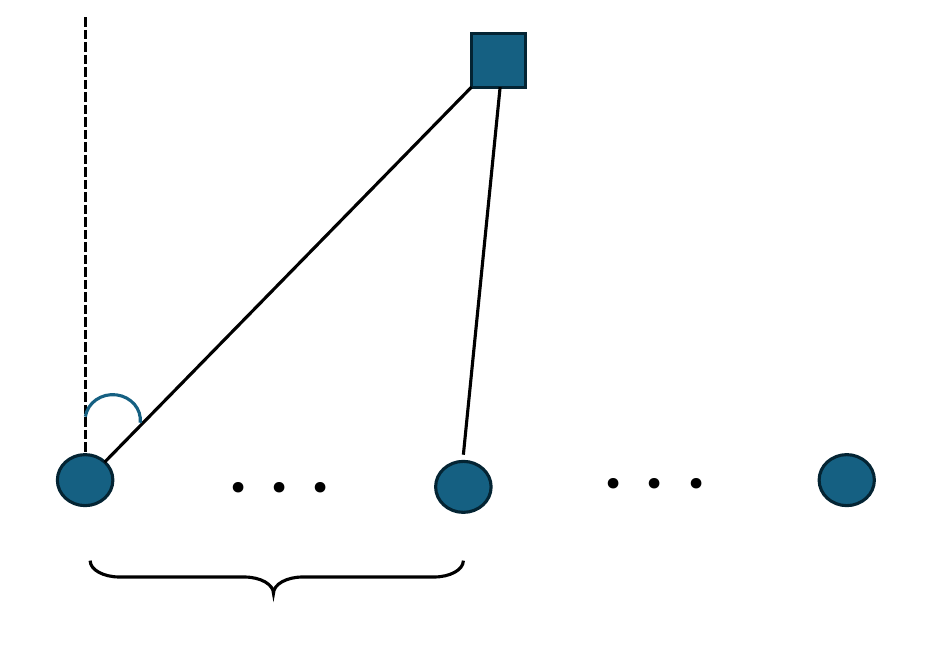}
  \put(11.9,28){$\theta_k$}%
  \put(18.9,0.1){$(m-1)d$}%
   \put(7.4,10.2){\footnotesize $1$}%
   \put(47.4,10.2){\footnotesize $m$}%
    \put(87,10.2){\footnotesize $M$}%
    \put(25,40){$r_k$}%
    \put(59,61){$k$th source}%
    \put(53,40){$\bar{r}_k^m(\theta)$}%
   \end{overpic}
\caption{ULA with $M$ antennas.}
\label{fig:antenna_array}
\end{figure}

Assume $K$ uncorrelated narrowband signals with unknown DoAs, $\theta_1,\theta_2,\ldots,\theta_K$, impinge on a uniform linear array (ULA) with $M$ antennas. The sources have a line-of-sight connection to the receiving array and they are assumed to be located in the far-field. The antenna separation is denoted by $d$. We assume $K<M$, which will later be required by the MUSIC algorithm to construct a noise subspace. At time slot $t$, the received signal by the antenna array is obtained as 
\begin{equation}
\label{eq:received_signal}
    \vect{y}(t) = \sum_{k=1}^K e^{j\psi_k}\vect{a}(\theta_k) s_k(t) + \vect{n}(t),
\end{equation}
where $\vect{y}(t) = [y_1(t),\ldots, y_{M}(t)]^\T$ is the received signal at the antenna array, $s_k(t)$ is the random signal from the $k$th source, which is unknown at the receiver. The $\vect{n}(t) = [n_1(t),\ldots,n_M(t)]^\T\sim \CN(\vect{0},\sigma^2\vect{I}_M)$ is the additive independent complex Gaussian noise and the term $e^{j\psi_k}$ in \eqref{eq:received_signal} represents the phase-shift on the first antenna of the array. The $\vect{a}(\theta_k)$ is the array response vector for the $k$th source, expressed as 
\begin{equation}
\label{eq:array_response_general}
\vect{a}(\theta_k) =  \left[1,\ldots,e^{-j\frac{2\pi}{\lambda}\left( \bar{r}_k^m (\theta_k)-r_k\right)},\ldots, e^{-j\frac{2\pi}{\lambda}\left( \bar{r}_k^M (\theta_k)-r_k\right)}\right]^\T,  
\end{equation}
where $r_k$ is the distance between the $k$th source and the first antenna, and $\bar{r}_k^m(\theta_k)$ denotes the distance between the source and the $m$th antenna ($m \neq 1$). The $\lambda$ represents the wavelength of the transmitted signals. Consider the antenna array shown in Figure~\ref{fig:antenna_array}. According to the figure, $\bar{r}_k^m(\theta_k)$ is obtained as 
\begin{align}
\label{eq:distance_from_antenna_m}
    \bar{r}_k^m (\theta_k)&= \sqrt{r_k^2 + (m-1)^2 d^2 - 2r_k(m-1)d \sin(\theta_k)} \nonumber\\
    &= r_k\sqrt{1 + \frac{(m-1)^2d^2}{r_k^2} - 2\frac{(m-1)d}{r_k}\sin(\theta_k)} \nonumber \\
    &\overset{(a)}{\approx} r_k \left( 1 + \frac{(m-1)^2d^2}{2r_k^2} - \frac{(m-1)d}{r_k}\sin(\theta_k)\right)\nonumber\\
    &\overset{(b)}{\approx} r_k - (m-1)d\sin(\theta_k) ,
    \end{align}
    where the first order Taylor approximation $\sqrt{1+ x } \approx 1 + 
 \frac{x}{2}$ for $|x|\ll 1$ is used in $(a)$, and $(b)$ holds because $d^2 \ll r_k^2$ in the far field. Substituting \eqref{eq:distance_from_antenna_m} into \eqref{eq:array_response_general}, we arrive at 
\begin{equation}
\label{eq:array_response}
  \vect{a}(\theta_k) = \left[1,\ldots,e^{j(m-1)\gamma_k  },\ldots,e^{j(M-1)\gamma_k } \right]^\T,  
\end{equation}where $\gamma_k =  2\pi \frac{d}{\lambda}\sin(\theta_k)$. 

Collecting all array response vectors in matrix $\vect{A}=\left[\vect{a}(\theta_1),\ldots,\vect{a}(\theta_K) \right] \in \mathbb{C}^{M \times K}$ and all source signals in vector $\vect{s}= [s_1(t),\ldots,s_K(t)]^\T \in \mathbb{C}^K$, \eqref{eq:received_signal} can be re-written as 
\begin{equation}
\label{eq:received_signal_vector}
 \vect{y}(t) = \vect{A}\diag\left(e^{j\psi_1},\ldots,e^{j\psi_K}\right)\vect{s}(t) + \vect{n}(t).
\end{equation}
The covariance matrix of the received signal is given by 
\begin{equation}
  \vect{R} = \mathbb{E}\left\{\vect{y}(t) \vect{y}^{\H} (t)\right\} = \vect{A} \vect{S} \vect{A}^\H + \sigma^2 \vect{I}_M,  \label{eq:R-ASA}  
\end{equation}where $\vect{S} = \mathbb{E}\left\{\vect{s}(t) \vect{s}^\H(t) \right\}$ is the source covariance matrix. The eigenvalue decomposition of the positive semi-definite Hermitian matrix $\vect{A}\vect{S}\vect{A}^\H$ in \eqref{eq:R-ASA} can be written as
\begin{align}
\vect{A}\vect{S}\vect{A}^\H = \vect{U}\dot{\bl\Sigma}\vect{U}^\H,
\end{align}
where the diagonal entries of $\dot{\bl\Sigma}$ contain the eigenvalues, which are real-valued and positive, arranged in descending order. The columns of $\vect{U}$ correspond to the unit-length eigenvectors. By augmenting a scaled identity matrix, as shown in \eqref{eq:R-ASA}, to the product of $\vect{U}\dot{\bl\Sigma}\vect{U}^\H$, the eigenvectors remain unaltered, while the eigenvalues are increased by $\sigma^2$. Hence, the eigenvalue decomposition of $\vect{R}$ yields 
\begin{equation}
    \vect{R} = \vect{A}\vect{S}\vect{A}^\H+\sigma^2\vect{I}_M = \vect{U}\underbrace{\left(\dot{\bl\Sigma}+\sigma^2\vect{I}_M\right)}_{=\bl{\Sigma}}\vect{U}^\H=\vect{U}_s \bl{\Sigma}_s \vect{U}_s^\H + \vect{U}_n \bl{\Sigma}_n \vect{U}_n^\H \label{eq:R-eigendec}
\end{equation}where $\bl{\Sigma}_s \in \mathbb{C}^{r \times r}$  and $\bl{\Sigma}_n \in \mathbb{C}^{(M-r) \times (M-r)}$ are diagonal matrices having the $r$ largest and $(M- r)$ smallest eigenvalues of $\vect{R}$ on their diagonal, respectively. Here, $r$ is the rank of the matrix $\vect{A}\vect{S}\vect{A}^\H$, and, thus, the $r$ largest eigenvalues on the diagonal of $\bl{\Sigma}_s$ are strictly greater than $\sigma^2$. On the other hand, the $M-r$ eigenvalues on the diagonal of $\bl{\Sigma}_n$ are exactly equal to $\sigma^2$. 
Matrices $\vect{U}_s \in \mathbb{C}^{M \times r}$ and $\vect{U}_n \in \mathbb{C}^{M \times (M-r)}$ contain the corresponding eigenvectors. Specifically, the columns of $\vect{U}_n$ span the noise subspace of $\vect{R}$ which is orthogonal its signal subspace. This property will be later used for estimating the directions of the sources.

The rank $r$ of $\vect{A}\vect{S}\vect{A}^\H$ is at most $K$ and it achieves the maximum rank $K$ when the rank of $\vect{A}$ and $\vect{S}$ are both $K$. If the sources are not fully correlated (coherent), then $\vect{S}$ is not rank-deficient, and it has a rank of $K$. When $M>K$, as assumed before, and the condition in the following remark holds, the rank of $\vect{A}$ is $K$.
\begin{remark}
\label{rem:unambgious_array}
The rank of $\vect{A}$ is $K$ for a ULA with $d\leq \lambda/2$ if the $K$ DoAs result in distinctly different values of $\sin(\theta_k)$.
\end{remark}

When the columns of the array response matrix $\vect{A}$ are linearly independent, meaning that the requirement stated in Remark~\ref{rem:unambgious_array} is satisfied, the array is referred to be \emph{unambiguous}. This property allows for the estimation of angles that are unique \citep{Krim1996a}.

In practice, the theoretical signal covariance matrix is not available and it is estimated as the  sample average covariance matrix that, for $L$ samples, is obtained as \begin{equation}
   \hat{\vect{R}} = \frac{1}{L}  \sum_{t = 1}^L \vect{y}(t) \vect{y}^\H(t).  \label{eq:hatR}
\end{equation}

\subsection{MUSIC}
\label{sec:MUSIC}
 Due to the orthogonality between signal and noise subspaces, we have
\begin{equation}
   \vect{A}\vect{S}\vect{A}^\H \vect{U}_n = \vect{0}. \label{eq:ASAUn}
\end{equation}
From linear algebra, we know that the rank of $\vect{A}\vect{S}\in\mathbb{C}^{M \times K}$ is equal to the rank of $\vect{A}$ when the source signal covariance matrix $\vect{S}$ is non-singular under the assumption $K<M$. Moreover, if the condition in Remark~\ref{rem:unambgious_array} holds, then $\vect{A}\vect{S}$ has the maximum rank of $K$. Then, \eqref{eq:ASAUn} implies
\begin{align}
 \vect{A}^\H\vect{U}_{n} =\vect{0} \quad &  \Rightarrow  \quad\vect{a}^\H(\theta_k)\vect{U}_n = \vect{0}, \quad k=1,\ldots,K \nonumber \\
 & \Rightarrow  \quad \vect{a}^\H(\theta_k)\vect{U}_n\vect{U}_{n}^\H\vect{a}(\theta_k) = 0, \quad k=1,\ldots,K. \label{eq:APA-trueR-condition}
\end{align}

If we have knowledge of $\vect{U}_{n}$ and the array is unambiguous, we may determine the DoA angles of the sources by identifying $K$ array response vectors that are linearly independent and satisfy the requirement stated in \eqref{eq:APA-trueR-condition}. The MUSIC algorithm is based on this premise, but it specifically addresses the scenario where the estimation of $\vect{U}_{n}$ is derived from the sample average covariance matrix $\hat{\vect{R}}$ in \eqref{eq:hatR}.

After computing the sample average covariance matrix, we compute its eigenvalue decomposition and obtain the noise subspace matrix $\hat{\vect{U}}_{n}\in \mathbb{C}^{M \times (M-K)}$ whose columns are the unit-length eigenvectors corresponding to the $M-K$ smallest eigenvalues of $\hat{\vect{R}}$. Using  the fact that $ \vect{a}^\H(\theta)\vect{U}_{n}\vect{U}_{n}^\H\vect{a}(\theta) = 0$ when considering the DoA of a source, we define the MUSIC spectrum as
\begin{align} \label{eq:MUSIC-spectrum}
P_{\textrm{MUSIC}}^{\,\textrm{far}}(\theta) = \frac{1}{\vect{a}^\H(\theta)\hat{\vect{U}}_{ n}\hat{\vect{U}}_{n}^\H\vect{a}(\theta)}
\end{align}
for angles $\theta \in \left[-\frac{\pi}{2},\frac{\pi}{2}\right]$ (assuming the signals impinge on the array by the front side). When the angle $\theta$ is in close proximity to a source, the denominator approaches zero, resulting in a peak in the spectrum.
If $\hat{\vect{U}}_{n}$ were exactly equal to $\vect{U}_{ n}$ (that is, $\hat{\vect{R}}=\vect{R}$), then the MUSIC spectrum would be infinite in true DoAs.
As we only possess the estimate $\hat{\vect{U}}_{n}$, the peak values and locations are mere approximations. The DoA estimations are determined by identifying the $K$ highest peaks in the MUSIC spectrum, given that the value of $K$ is already known. The number of sources can be determined using the Akaike information or the minimum description length criterion \citep{Wax1985}. Furthermore, the MUSIC algorithm may detect the number of sources by determining the number of eigenvalues of the estimated covariance matrix, $\hat{\vect{R}}$,  that significantly exceed $\sigma^2$.  Utilizing this quantity as the estimation for $K$, we proceed by determining the $K$ highest peaks of the MUSIC spectrum.

\begin{figure}
    \centering
    \includegraphics[width = 0.88\columnwidth]{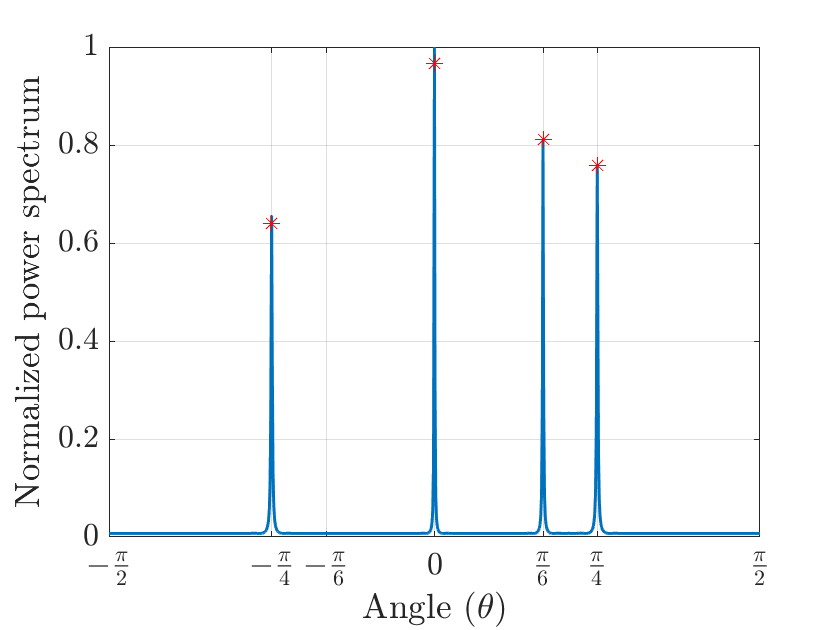}
    \caption{The normalized power spectrum for the MUSIC algorithm when the number of antennas is $M=50$. The source DoAs are given $-\pi/4$, $0$, $\pi/6$, and $\pi/4$, respectively, which are shown by red stars.}
    \label{fig:FF-MUSIC}
\end{figure}

In Figure~\ref{fig:FF-MUSIC}, we plot the MUSIC spectrum for a scenario with $\lambda/2$-spacing antenna array with $M=50$ antennas. There are $K=4$ sources with independent data streams, that are zero-mean complex Gaussian distributed with unit variance, and equal SNR of $0$\,dB per antenna and per sample. Hence, the source covariance matrix $\vect{S}$ is a scaled identity matrix. The DoAs of the sources are given by $-\pi/4$, $0$, $\pi/6$, and $\pi/4$, respectively. The number of samples is $L=100$. As demonstrated in the figure,  the correct DoA angles can be almost identified from the four highest peaks of the MUSIC spectrum.

\begin{figure}
	\centering
	\begin{subfigure}{\columnwidth}
		\centering
		\includegraphics[width=0.88\columnwidth]{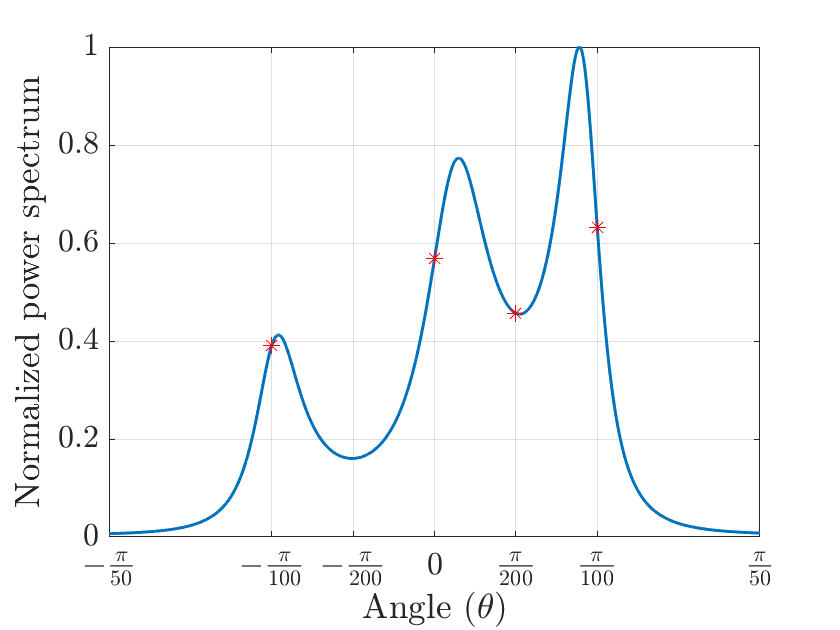}
		\caption{$M=50$ antennas}
		\label{fig:FF-MUSIC-closely-50}
  \vspace{-1mm}
	\end{subfigure}
	\begin{subfigure}{\columnwidth}
		\centering
		\includegraphics[width=0.88\columnwidth]{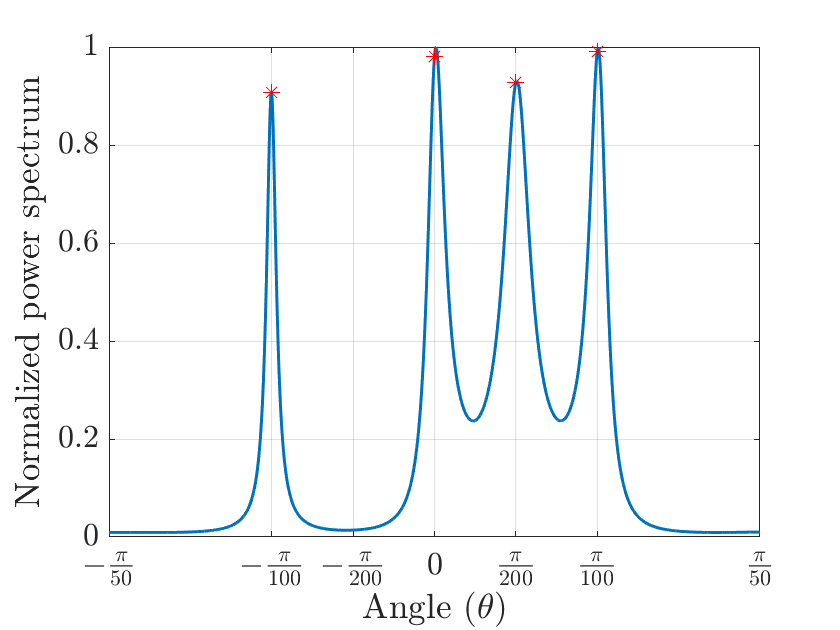}
		\caption{$M=100$ antennas}
		\label{fig:FF-MUSIC-closely-100}
  \vspace{-3mm}
	\end{subfigure}
 \caption{The normalized power spectra of the MUSIC algorithm when the four source angles are very close. The DoAs are $-\pi/100$, $0$, $\pi/200$, and $\pi/100$, respectively, which are shown by red stars.  }
 \label{fig:FF-MUSIC-closely}
\end{figure}

To gain a more comprehensive understanding of the limitations of the MUSIC method in accurately calculating the DoA angles, we will now position the four sources at the specific angles of $-\pi/100$, $0$, $\pi/200$, and $\pi/100$ accordingly. Figure~\ref{fig:FF-MUSIC-closely}(a) displays three peaks; moreover, their angles do not correspond to the correct angles. Due to the near proximity of the sources in the angular domain, the resolution offered by a set of $M=50$ antennas is inadequate. This is because one of the two nearby angles falls inside the beamwidth of a beam pointing at the other DoA. In this case, the two peaks merge. To alleviate this issue, the beamwidth can be made narrower by increasing the number of antennas. In Figure~\ref{fig:FF-MUSIC-closely}(b), we increase the number of antennas to $M=100$ while maintaining the same configuration. This time, it is seen that the correct angles can be identified thanks to the increased spatial resolution by doubling the antennas.

\subsection{ESPRIT}
\label{sec:ESPRIT}

The ESPRIT algorithm differs from the MUSIC algorithm in utilizing the signal subspace instead of the noise subspace to estimate the DoA angles. In order to explain how the angles are determined using the signal subspace estimate obtained from the sample average covariance matrix in \eqref{eq:hatR}, we will provide an overview of the theoretical framework of the ESPRIT method. We assume $K\leq M$ (unlike MUSIC algorithm, ESPRIT can work when $K=M$) and that $\vect{A}\vect{S}\vect{A}^\H$ has the maximum rank of $K$, i.e., the dimension of the signal subspace is $K$.

We first multiply equation \eqref{eq:R-eigendec} from the right by $\vect{U}_s$ and obtain
\begin{align}
&\vect{R}\vect{U}_s = \vect{A}\vect{S}\vect{A}^\H\vect{U}_s+\sigma^2\vect{U}_s =\vect{U}_s \bl{\Sigma}_s \underbrace{\vect{U}_s^\H\vect{U}_s}_{=\vect{I}_K} + \vect{U}_n \bl{\Sigma}_n \underbrace{\vect{U}_n^\H\vect{U}_s}_{=\vect{0}}=\vect{U}_s\bl{\Sigma}_s \nonumber\\
& \Rightarrow  \quad  \vect{A}\vect{S}\vect{A}^\H\vect{U}_s = \vect{U}_s\underbrace{\left(\bl{\Sigma}_s-\sigma^2\vect{I}_K\right)}_{=\overline{\bl{\Sigma}}_s} \label{eq:Us-mult},
\end{align}
where the diagonal entries of $\overline{\bl{\Sigma}}_s$ are strictly greater than zero since the signal space eigenvalues appearing on the diagonal of $\bl{\Sigma}_s$ are strictly greater than $\sigma^2$. Hence, $\overline{\bl{\Sigma}}_s$ is non-singular. Multiplying the both sides of \eqref{eq:Us-mult} from the right by $\overline{\bl{\Sigma}}_s^{-1}$ we obtain
\begin{align}
    \vect{U}_s = \vect{A}\underbrace{\vect{S}\vect{A}^\H\vect{U}_s\overline{\bl{\Sigma}}_s^{-1}}_{=\vect{C}}=\vect{A}\vect{C}. \label{eq:Us-AC}
\end{align}
We will now partition ULA into two overlapping subarrays, one containing the antennas $1$ to $M-1$ and the other having antennas $2$ to $M$. Let $\vect{A}_1$ and $\vect{A}_2$ be their corresponding array response matrices
such that
\begin{equation}
\label{eq:array_response_split0}
    \vect{A}=
\begin{bmatrix}
\vect{A}_1 \\
 \mathrm{Last~row}
\end{bmatrix} = \begin{bmatrix}
\mathrm{First~row} \\
\vect{A}_2
\end{bmatrix}. 
\end{equation}
Using the symmetric structure of the ULA, the array response matrix of the second subarray can be represented as 
\begin{equation}
   \vect{A}_2 =  \vect{A}_1\vect{D}, \label{eq:A2DA1}
\end{equation}where 
\begin{equation}
    \vect{D}=
\begin{bmatrix}
e^{j \frac{2\pi d}{\lambda}\sin(\theta_1) } &  & 0 \\
 & \ddots &  \\
0 &  & e^{j \frac{2\pi d}{\lambda}\sin (\theta_K) }.
\end{bmatrix} \label{eq:D}
\end{equation}
Similar to \eqref{eq:array_response_split0}, we can split $\vect{U}_s=\vect{AC}$ from \eqref{eq:Us-AC} as 
\begin{align}
\vect{U}_s &=
\begin{bmatrix}
\vect{U}_{s1} \\
 \mathrm{Last~row} 
\end{bmatrix} = \begin{bmatrix}
\mathrm{First~row} \\
 \vect{U}_{s2}
\end{bmatrix}, \\
\vect{AC} &=
\begin{bmatrix}
\vect{A}_1 \vect{C} \\
 \mathrm{Last~row}
\end{bmatrix} = \begin{bmatrix}
\mathrm{First~row} \\
 \vect{A}_2 \vect{C}
\end{bmatrix},
\end{align}which implies that 
\begin{align}
   \vect{U}_{s1} &= \vect{A}_1 \vect{C}, \label{eq:Us1}\\ 
 \vect{U}_{s2} &= \vect{A}_2 \vect{C}. \label{eq:Us2}
\end{align}
Substituting $\vect{A}_2=\vect{A}_1\vect{D}$ into \eqref{eq:Us2}, we obtain
\begin{align}
 \vect{U}_{s2}  = \vect{A}_1\vect{D}\vect{C}.
\end{align}
Since the ranks of $\vect{S}$, $\vect{A}$, $\vect{U}_s$, and $\overline{\bl{\Sigma}}_s$ are all $K$, the rank of $K\times K$ matrix $\vect{C}$ is also  $K$. Hence, $\vect{C}$ is invertible, and we have the relation $\vect{A}_1=\vect{U}_{s1}\vect{C}^{-1}$ from \eqref{eq:Us1}. Inserting this into the above expression, we end up with
\begin{align}
    \vect{U}_{s2} =\vect{U}_{s1}\underbrace{\vect{C}^{-1}\vect{D}\vect{C}}_{\bl{\Phi}},
\end{align}
where the matrix $\bl{\Phi}$ has the same eigenvalues as $\vect{D}$, which are the diagonal entries given in \eqref{eq:D}. This is due to the fact that $\bl{\Phi}$ is obtained through a \emph{similarity transformation} from $\vect{D}$ using the matrix $\vect{C}$. Multiplying both sides of the above equation from the left by the left pseudo-inverse of $\vect{U}_{s1}$, i.e., $\left(\vect{U}_{s1}^\H\vect{U}_{s1}\right)^{-1}\vect{U}_{s1}^\H$, we arrive at
\begin{align}
    \bl{\Phi} = \left(\vect{U}_{s1}^\H\vect{U}_{s1}\right)^{-1}\vect{U}_{s1}^\H\vect{U}_{s2}
\end{align}
whose eigenvalues give the functions $e^{j\frac{2\pi d}{\lambda}\sin(\theta_k)}$ of the DoA angles.

Now, from the eigenvalue decomposition of the sample average covariance matrix $\hat{\vect{R}}$ in \eqref{eq:hatR}, we can obtain $\hat{\vect{U}}_{s1}\in \mathbb{C}^{(M-1)\times K}$ and $\hat{\vect{U}}_{s2}\in \mathbb{C}^{(M-1)\times K}$ and form the matrix
\begin{align}
    \hat{\bl{\Phi}} = \left(\hat{\vect{U}}_{s1}^\H\hat{\vect{U}}_{s1}\right)^{-1}\hat{\vect{U}}_{s1}^\H\hat{\vect{U}}_{s2}.
\end{align}Assuming that $\mu_1,\ldots,\mu_K$ are the eigenvalues of $\hat{\bl{\Phi}}$ and using $e^{j\frac{2\pi d}{\lambda}\sin(\theta_k)}$, the DoA estimates are obtained as
\begin{align}
    \hat{\theta}_k = \arcsin\left( \angle{\mu_k}\frac{\lambda}{2\pi d}\right), \quad k=1,\ldots,K,
\end{align}
where we have implicitly assumed $\lambda/d\leq2\Rightarrow d\geq\lambda/2$. For unambigious DoA estimate, we thus need to set $d=\lambda/2$.

In Tables~\ref{tab:ESPRIT-50} and \ref{tab:ESPRIT-100}, we compare the DoA estimates obtained with the MUSIC and ESPRIT algorithms for the setups in Figures~\ref{fig:FF-MUSIC} and \ref{fig:FF-MUSIC-closely}(b), respectively. It should be noted that the performance of the MUSIC algorithm depends on the size of the angular spectrum grid, which is selected as $100\,000$ points in this simulation. In contrast, ESPRIT is a grid-free algorithm. We observe that the performance of the MUSIC algorithm is slightly better compared to that of ESPRIT at the cost of increased complexity.

\begin{table}
\caption{Comparison of MUSIC and ESPRIT when there are $M=50$ antennas and the DoAs are $-\pi/4$, 0, $\pi/6$, and $\pi/4$.\label{tab:ESPRIT-50}}{%
\begin{tabular}{@{}ccccc@{}}
\toprule
\midrule
True DoA & $-0.7854$ & 0       & 0.5236 & 0.7854 \\
\midrule
MUSIC    & $-0.7859$ & $-0.0004$ & 0.5236 & 0.7856 \\
\midrule
ESPRIT   & $-0.7873$ & $-0.0002$ & 0.5242 & 0.7851 \\
\botrule
\end{tabular}}{}
\end{table}

\begin{table}
\caption{Comparison of MUSIC and ESPRIT when there are $M=100$ antennas and the DoAs are $-\pi/100$, $0$, $\pi/200$, and $\pi/100$.\label{tab:ESPRIT-100}}{%
\begin{tabular}{@{}ccccc@{}}
\toprule
\midrule
True DoA & $-0.0314$ & 0       & 0.0157 & 0.0314 \\
\midrule
MUSIC    & $-0.0315$ & 0.0003 & 0.0159 & 0.0316 \\
\midrule
ESPRIT   & $-0.0310$ & 0.0003 & 0.0150 & 0.0320 \\
\botrule
\end{tabular}}{}
\end{table}

\begin{figure}
    \centering
    \includegraphics[width = \columnwidth]{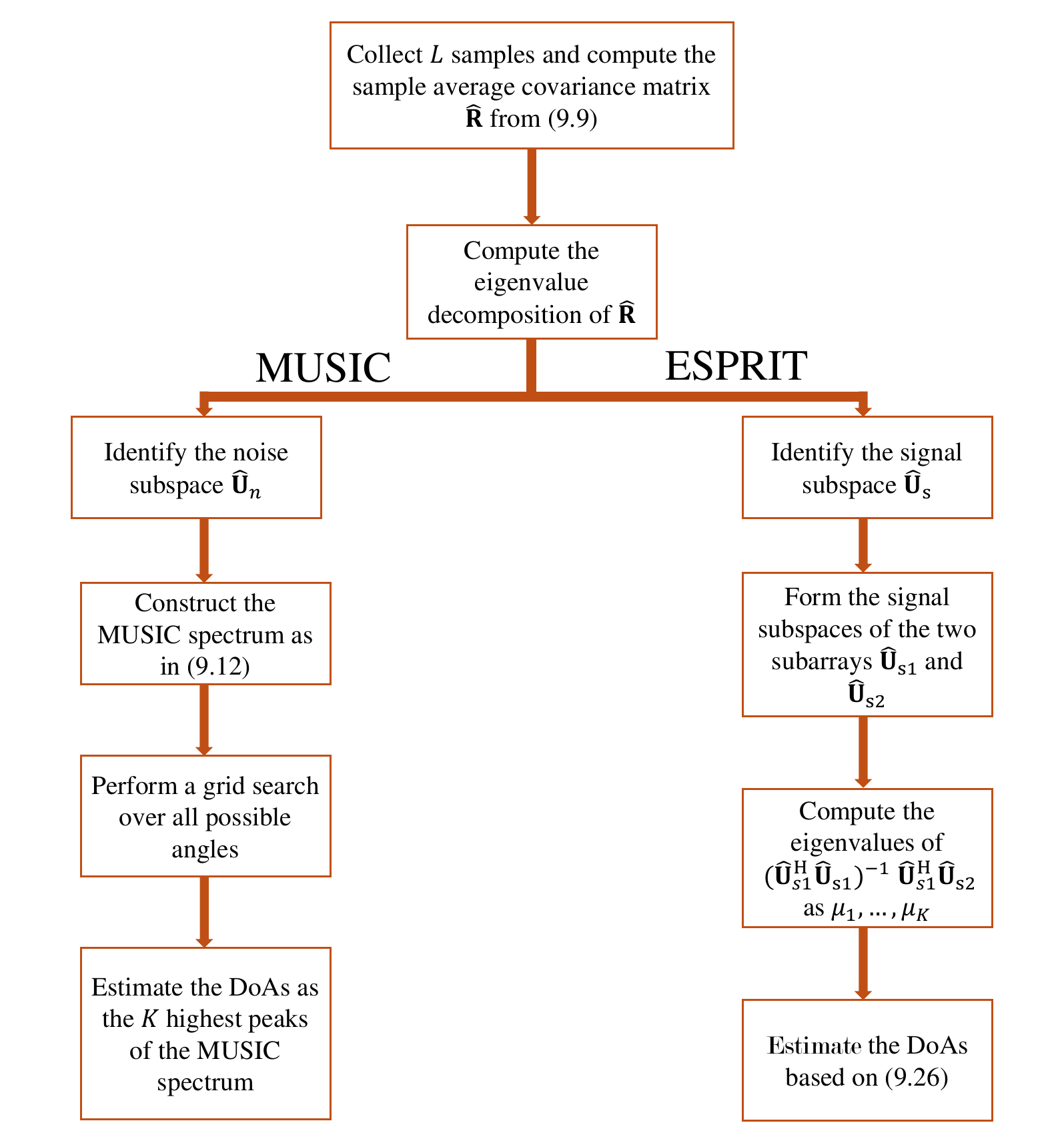}
    \caption{Flowchart of the MUSIC and ESPRIT algorithms for DoA estimation.}
    \label{fig:block_diagrams}
\end{figure}

Figure~\ref{fig:block_diagrams} shows the steps of DoA estimation using the MUSIC and ESPRIT algorithms.

\subsubsection{Generalized ESPRIT}
\label{sec:gen_ESPRIT}
The ESPRIT algorithm can be extended to a more general scenario, where the distance between two identically-indexed antennas in the subarrays does not need to be the same for all indices \citep{Gao2005generalized}. To clarify this, we consider an array with $M=2N$ antennas that is divided into two subarrays of $N$ antennas. It is assumed that $K \leq N$.
The antennas of each subarray can be arbitrarily chosen; therefore, the distance between the $n$th antenna in the first subarray and the corresponding antenna in the second subarray may be different for different values of $n$. The array response matrix $\vect{A}$ can be expressed as 

\begin{equation}
\label{eq:A_gen_ESPRIT}
   \vect{A} = \begin{bmatrix}
       \vect{A}_1 \\
       \vect{A}_2
   \end{bmatrix}, 
\end{equation}where $\vect{A}_1$ and $\vect{A}_2$ are the array response matrices of the two subarrays. In particular, assuming that $\vect{a}_1(\theta)$ is the array response vector of the first subarray for a signal coming from the angle $\theta$, the array response matrices are given by   
\begin{align}
\vect{A}_1 &= [\vect{a}_1(\theta_1),\ldots,\vect{a}_1(\theta_K)]\in \mathbb{C}^{N\times K},    \\
\vect{A}_2 &= [\bl{\Lambda}(\theta_1) \vect{a}_1(\theta_1),\ldots,\bl{\Lambda}(\theta_K) \vect{a}_1(\theta_K)] \in \mathbb{C}^{N\times K}, 
\end{align}where 
\begin{equation}
    \bl{\Lambda}(\theta_k)= 
\begin{bmatrix}
e^{j \frac{2\pi }{\lambda} l_1 \sin (\theta_k)} &  & 0 \\
 & \ddots &  \\
0 &  & e^{j \frac{2\pi }{\lambda} l_N \sin(\theta_k)}
\end{bmatrix}
\end{equation}and $l_n$ denotes the distance between the $n$th antennas of the second and first subarrays. Similarly to \eqref{eq:A_gen_ESPRIT}, $\vect{U}_s = \vect{AC}$ can be written as 
\begin{equation}
   \vect{U}_s = \begin{bmatrix}
       \vect{U}_{s1} \\
       \vect{U}_{s2}
   \end{bmatrix} = \begin{bmatrix}
       \vect{A}_1 \vect{C} \\
       \vect{A}_2 \vect{C}
   \end{bmatrix}.
\end{equation}Introducing the diagonal matrix 
\begin{equation}
    \bl{\Psi}(\theta)= 
\begin{bmatrix}
e^{j \frac{2\pi }{\lambda} l_1 \sin (\theta)} &  & 0 \\
 & \ddots &  \\
0 &  & e^{j \frac{2\pi }{\lambda} l_N \sin ( \theta )}
\end{bmatrix},
\end{equation}we form the matrix $\vect{F}(\theta) \in \mathbb{C}^{N \times K}$ as 
\begin{equation}
    \vect{F} (\theta) = \vect{U}_{s2} - \bl{\Psi}(\theta) \vect{U}_{s1} = \vect{Q}\vect{C},
\end{equation}where
\begin{align}
    \vect{Q} = \left[\left(\bl{\Lambda}(\theta_1) - \bl{\Psi}(\theta)\right)\vect{a}_1(\theta_1),\ldots, \left(\bl{\Lambda}(\theta_K) - \bl{\Psi}(\theta)\right)\vect{a}_1(\theta_K)\right].
\end{align}When $\theta = \theta_k$, the $k$th column of $\vect{Q}$ becomes zero and $\vect{F}(\theta)$ drops in rank. Therefore, matrix $\vect{W}^\H \vect{F}(\theta_k)$ is singular, where $\vect{W} \in \mathbb{C}^{N \times K}$ is an arbitrary full-rank matrix. The DoAs can then be estimated by finding the $K$ highest peaks of the following spectrum 
\begin{equation}
f(\theta) = \frac{1}{\det \left(\vect{W}^\H  \vect{F}(\theta) \right)}.
\end{equation}We will use the idea of generalized ESPRIT later for finding the DoAs of near-field sources. 

\section{Near-Field DoA and Range Estimation}
\label{sec:near-field}
This section considers the case where the sources are located in the radiative near-field region of the antenna array. The 2D MUSIC algorithm is first presented for simultaneously estimating the DoAs and ranges of the sources via a 2D search over the grid of possible angles and ranges. Then, two symmetry-based localization algorithms based on MUSIC and generalized ESPRIT are discussed that decompose DoA and range estimation problems and reduce the search complexity by turning a 2D search into multiple 1D searches.

\subsection{System Model}
 When the sources lie in the radiative near-field region, the signal wavefront is spherical and the range of the sources must be estimated along with their DoAs for accurate localization. Therefore, far-field DoA estimation methods based on the planar wavefront assumption cannot be used directly for source localization in the radiative near-field region.  Denoting by $r_1, r_2,\ldots,r_K$ the range of the sources, which are located in the far-field of a single antenna but in the radiative near-field of the whole array, we have
\begin{equation}
   r_k \in \left[ d_\F, d_\FA \right ], ~ k = 1,2,\ldots,K,  
\end{equation}where $d_\F$ is the Fraunhofer distance of a single antenna given by 
\begin{equation}
  d_\F = \frac{2D^2}{\lambda},  
\end{equation}with $D$ being the antenna aperture length. Likewise, the Fraunhofer array distance for an antenna array can be defined as \citep{ramezani2022bookchapter}
\begin{equation}
    d_\FA = \frac{2W^2}{\lambda},
\end{equation}where $W$ is the array aperture length. For an aperture ULA of $M$ antennas, the Fraunhofer array distance can be approximated as \citep{ramezani2023magazine}
\begin{equation}
   d_\FA \approx \frac{M^2}{2}\,d_\F.
\end{equation}
Similar to the far-field scenario, the received signal at the antenna array can be expressed as 
\begin{equation}
\label{eq:received_signal_vector_near-field}
 \vect{y}(t) =  \vect{A}\diag\left(e^{j\psi_1},\ldots,e^{j\psi_K}\right)\vect{s}(t) + \vect{n}(t),   
\end{equation}where the array response matrix in the near-field is given by 
\begin{equation}
    \vect{A} = \left[\vect{a}(\theta_1,r_1),\ldots,\vect{a}(\theta_K,r_K) \right] \in \mathbb{C}^{M \times K},
\end{equation}and the near-field array response vector for the signal coming from the $k$th source is  
\begin{equation}
\label{eq:array_response_general2}
\vect{a}(\theta_k,r_k) =  \left[1,\ldots,e^{-j\frac{2\pi}{\lambda}\left(\bar{r}_k^m (\theta_k,r_k)-r_k\right)},\ldots, e^{-j\frac{2\pi}{\lambda}\left( \bar{r}_k^M (\theta_k,r_k)-r_k\right)}\right]^\T,  
\end{equation}where the distance between the $k$th source and $m$th antenna, $\bar{r}_k^m (\theta_k,r_k)$ is a function of both the DoA and range of the source. Specifically, 
\begin{align}
\label{eq:distance_from_antenna_m_second}
    \bar{r}_k^m (\theta_k)&= \sqrt{r_k^2 + (m-1)^2 d^2 - 2r_k(m-1)d \sin(\theta_k)}  \nonumber \\
    &\overset{(a)}{\approx} r_k \Bigg( 1 + \frac{(m-1)^2d^2}{2r_k^2} - \frac{(m-1)d}{r_k}\sin(\theta_k)- \frac{(m-1)^4d^4}{8 r_k^4}\nonumber\\
    &\quad \quad \quad+ \frac{(m-1)^3d^3}{2r_k^3}\sin(\theta_k) - \frac{(m-1)^2d^2}{2r_k^2}\sin^2(\theta_k) \Bigg) \nonumber \\
    & \overset{(b)}{\approx} r_k \left(1 - \frac{(m-1)d}{r_k}\sin(\theta_k) + \frac{(m-1)^2d^2}{2r_k^2}\left(1 - \sin^2 (\theta_k)\right) \right),
    \end{align}
where the second order Taylor approximation $\sqrt{1+x} \approx 1 + \frac{x}{2} - \frac{x^2}{8}$, for $|x|\ll 1$ is used in $(a)$ and $(b)$ is obtained by assuming $d^3 \ll r_k^3$ and $d^4 \ll r_k^4$. Therefore, the near-field array response vector can be expressed as  
\begin{equation}
\label{eq:array_response_near-field}
  \vect{a}(\theta_k,r_k) = \left[1,\ldots,e^{j\left((m-1)\gamma_k + (m-1)^2 \phi_k\right) },\ldots,e^{j\left((M-1)\gamma_k + (M-1)^2 \phi_k\right) } \right]^\T, 
\end{equation}where $\phi_k =- \pi \frac{d^2}{\lambda r_k} \cos^2(\theta_k)$.

\begin{remark}
   According to \eqref{eq:array_response_near-field}, the array is said to be unambiguous if $d \leq \lambda/2$. In such a case, the entries of the array response vector will be distinct for different values of $\theta_k$. 
\end{remark}

Recall from Section~\ref{sec:far-field} that the covariance matrix of the received signal is obtained as 
\begin{equation}
  \vect{R} = \mathbb{E}\left\{\vect{y}(t) \vect{y}^{\H} (t)\right\} = \vect{A} \vect{S} \vect{A}^\H + \sigma^2 \vect{I}_M =  \vect{U}_s \bl{\Sigma}_s \vect{U}_s^\H + \vect{U}_n \bl{\Sigma}_n \vect{U}_n^\H, 
\end{equation}and the sample average covariance matrix using $L$ samples is constructed as 
\begin{equation}
   \hat{\vect{R}} = \frac{1}{L}  \sum_{t = 1}^L \vect{y}(t) \vect{y}^\H(t). 
   \end{equation}

\subsection{Two-Dimensional MUSIC}
2D MUSIC extends 1D MUSIC by performing the grid search over two dimensions, i.e., DoA and range.
Similar to the 1D MUSIC method described in Section~\ref{sec:MUSIC}, 2D MUSIC exploits the orthogonality between the signal and noise subspaces for estimating the location of the sources. According to \eqref{eq:APA-trueR-condition}, we have
\begin{figure}
	\centering
	\begin{subfigure}{\columnwidth}
		\centering
		\includegraphics[width=0.9\columnwidth]{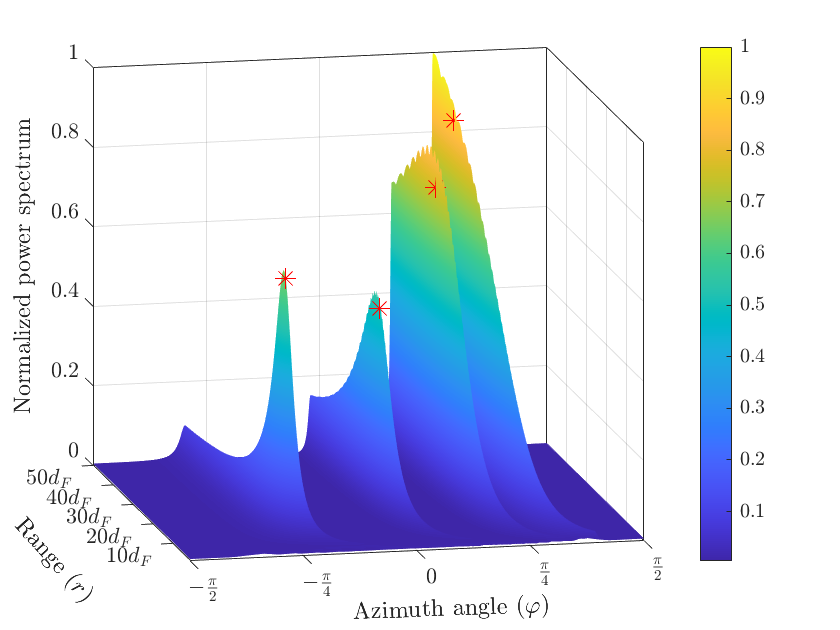}
		\caption{$M = 10$}
		\label{fig:MUSIC10}
	\end{subfigure}
	\begin{subfigure}{\columnwidth}
		\centering
		\includegraphics[width=0.9\columnwidth]{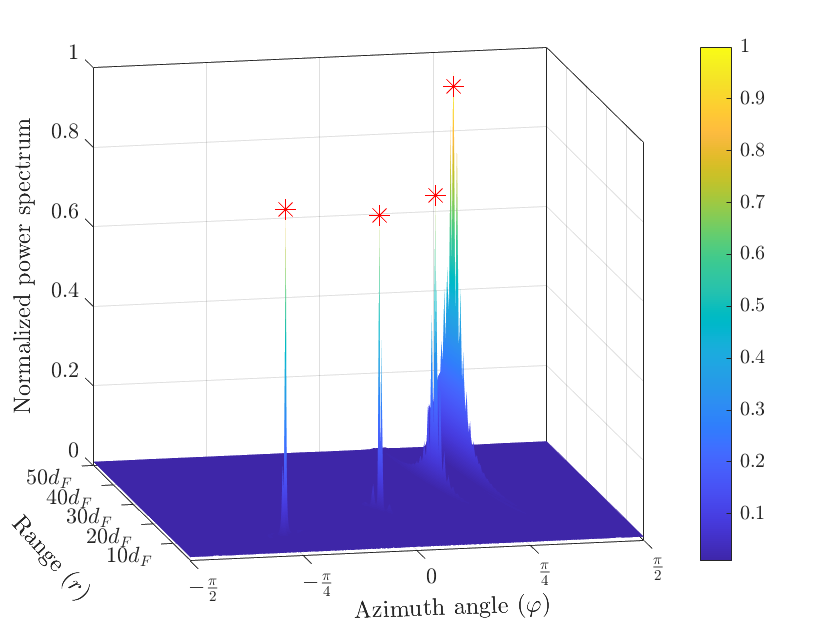}
		\caption{$M=50$}
		\label{fig:MUSIC50}
	\end{subfigure}
 \caption{The normalized power spectra for the 2D MUSIC when there are four sources located at $(-\pi/4,10\,d_\F)$, $(0,20\,d_\F)$, $(\pi/6,30\,d_\F)$, and $(\pi/4,40\,d_\F)$, shown by red stars on the figures. }
 \label{fig:2D_MUSIC}
\end{figure}
\begin{align}
    \vect{A}^\H\vect{U}_n = \vect{0} \quad &\Rightarrow \quad \vect{a}^\H(\theta_k,r_k) \vect{U}_n= \vect{0}, \quad k=1,\ldots,K \nonumber \\
    &\Rightarrow \quad \vect{a}^\H(\theta_k,r_k)\vect{U}_n \vect{U}_n^\H \vect{a}(\theta_k,r_k) = 0, \quad k = 1,\ldots, K.
\end{align}
Using the estimate $\hat{\vect{U}}_n$ that contains the eigenvectors corresponding to the $M-K$ smallest eigenvalues of $\hat{\vect{R}}$, the angles and ranges of the sources can be obtained by identifying the $K$ highest peaks of the spectrum 
\begin{align} \label{eq:2D_MUSIC-spectrum}
P_{\textrm{MUSIC}}^{\,\textrm{near}}(\theta,r) = \frac{1}{\vect{a}^\H(\theta,r)\hat{\vect{U}}_{ n}\hat{\vect{U}}_{n}^\H\vect{a}(\theta,r)},
\end{align}which can be done via a 2D grid search over angles $\theta \in \left[-\frac{\pi}{2},\frac{\pi}{2}\right]$ and ranges $r \in [d_\F, d_\FA]$.  

Figure~\ref{fig:2D_MUSIC} evaluates the performance of the 2D MUSIC when there are $K = 4$ near-field sources to be localized. Two cases for number of antennas at the array is considered: $M = 10$ and $M = 50$. The spacing between the antennas is assumed to be $d = \lambda/2$, the number of samples is set as $L = 100$, and the SNR is $0\,$dB per antenna and per sample. The angle-range pairs of the sources are given by $(-\pi/4,10\,d_\F)$, $(0,20\,d_\F)$, $(\pi/6,30\,d_\F)$, and $(\pi/4,40\,d_\F)$. The figures demonstrate the power spectrum in \eqref{eq:2D_MUSIC-spectrum} over different angles and ranges, when the peak value of the spectrum is normalized to one. The red stars show the locations of the sources. In case of $M=10$ (Figure~\ref{fig:2D_MUSIC}(a)), the peaks of the spectrum do not exactly match with the location of the sources. When we increase the number of antennas to $M=50$, the algorithm can estimate the locations more accurately due to the higher spatial resolution in the power spectrum, as depicted in Figure~\ref{fig:2D_MUSIC}(b). 
    
\subsection{Symmetry-Based Localization}
The MUSIC algorithm incurs high complexity for localization of near-field sources due to performing a 2D grid search over all possible angles and ranges.
Decomposing the localization problem into separate DoA and range estimation problems can remarkably reduce the search complexity. However, the DoAs and ranges of the sources are coupled in the expression of the array response vector in \eqref{eq:array_response_near-field}, which means that the two problems cannot be directly separated. To deal with this issue, some methods have been proposed that, utilizing the symmetry property of the array, decouple the DoA and range estimation problems and solve the localization problem through multiple 1D grid searches
\citep{Zhi2007nearfield,He2012efficient,Liu2013Efficient}. The strategy is to first find the DoAs of the $K$ sources by forming a spectrum function that can be utilized for identifying the DoAs independently of the ranges. The ranges are then obtained by performing the standard MUSIC algorithm once for each of the estimated DoAs, repeated $K$ times. Here, we will cover two methods, which decouple the DoA and range estimation problems by resorting to the symmetric structure of the antenna array. 

Consider a ULA consisting of $M = 2N+1$ antennas, where the antennas are indexed from $-N$ to $N$ and the antenna at the center of the array is considered as the reference antenna. With this setup, the array response vector in \eqref{eq:array_response_near-field} is modified as 
\begin{equation}
\label{eq:array_response_symmetric}
  \vect{a}(\theta_k,r_k) = \left[e^{j\left(-N\gamma_k + N^2 \phi_k\right) },\ldots,1,\ldots,e^{j\left(N\gamma_k + N^2 \phi_k\right) } \right]^\T.  
\end{equation}The received signal at this symmetric antenna array can be expressed as 
\begin{align}
   \vect{y}(t) &= \begin{bmatrix}
y_{-N}(t) \\
 \vdots \\
y_{N}(t)
\end{bmatrix}  \nonumber \\ &=  \begin{bmatrix}
e^{j\left(-N\gamma_1 + N^2 \phi_1\right)} & \ldots & e^{j\left(-N\gamma_K + N^2 \phi_K\right)} \\
\vdots & \ddots & \vdots  \\
e^{j\left(N\gamma_1 + N^2 \phi_1\right)} & \ldots & e^{j\left(N\gamma_K + N^2 \phi_K\right)} 
\end{bmatrix} \begin{bmatrix}
e^{j\psi_1}s_1(t) \\
 \vdots \\
e^{j\psi_K}s_K(t)
\end{bmatrix} + \begin{bmatrix}
n_{-N}(t) \\
 \vdots \\
n_{N}(t)
\end{bmatrix}. \label{eq:received_signal_symmetric}
\end{align}

 \subsubsection{Modified MUSIC for Near-Field Localization}
After computing the covariance matrix of the received signal $\vect{R} = \mathbb{E}\left\{\vect{y}(t) \vect{y}^{\H} (t)\right\} = \vect{A} \vect{S} \vect{A}^\H + \sigma^2 \vect{I}_M$, its anti-diagonal entries are given by 
\begin{align}
   \vect{R}[n,2N+2-n] &=\mathbb{E}\Bigg\{ \left( \sum_{k=1}^K e^{j\left((n-N-1)\gamma_k + (n-N-1)^2 \phi_k\right) }e^{j\psi_k} s_k(t)  \right) \nonumber\\
   &\quad\quad \times \left( \sum_{k=1}^K e^{-j\left((N-n+1)\gamma_k + (N-n+1)^2 \phi_k\right)} e^{-j\psi_k}s_k^*(t) \right) \Bigg\} \notag \\
   & \quad + \mathbb{E}\left\{ n_{n-N-1}(t) n^*_{N-n+1}(t) \right\} \notag \\ & = \sum_{k=1}^Kp_k e^{-j2(N-n+1)\gamma_k} + \sigma^2 \delta_{n,2N+2-n},\quad n = 1,\ldots,2N+1,
   \end{align}
   where $\vect{R}[i,j]$ denotes the entry in the $i$th row and $j$th column of $\vect{R}$, $\delta_{i,j}$ is the Kronecker delta function which returns $1$ when $i = j$ and $0$ otherwise. We have assumed that the source signals are independent and each has a zero-mean. The variance of $s_k(t)$ is denoted by $p_k$. We now construct a vector $\bar{\vect{y}} \in \mathbb{C}^{2N+1}$ where its $n$th entry is $\vect{R}[n,2N+2-n]$ without the noise variance, i.e, 
\begin{equation}
\label{eq:spatial_signature}
   \bar{\vect{y}} = \left[ \sum_{k=1}^Kp_k e^{-j2N \gamma_k},\sum_{k=1}^Kp_k e^{-j2(N-1) \gamma_k},\ldots,\sum_{k=1}^Kp_k e^{j2(N-1) \gamma_k},\sum_{k=1}^Kp_k e^{j2N \gamma_k}\right]^\T. 
\end{equation} The entries of $\bar{\vect{y}}$ only contain the angle information, which signifies that DoA estimation can be performed independently of the range estimation. To this end, $\bar{\vect{y}}$ is split into $J$ overlapping subvectors, each containing $2N+2-J$ entries. Specifically, the $i$th subvector is formed as 
\begin{equation}
  \bar{\vect{y}}_i = \left[\sum_{k=1}^Kp_k e^{-j2(N-i+1) \gamma_k},\ldots, \sum_{k=1}^Kp_k e^{-j2(J-i-N) \gamma_k} \right],  
\end{equation}which can be decomposed as 
\begin{equation}
   \bar{\vect{y}}_i = \vect{B}\vect{p}_i, 
\end{equation}where
\begin{align}
   &\vect{B} = [\vect{b}(\gamma_1),\ldots,\vect{b}(\gamma_K)]\in \mathbb{C}^{(2N+2-J) \times K}, \\
   &\vect{p}_i = \left[p_1e^{j2i \gamma_1},\ldots,p_Ke^{j2i\gamma_K} \right]^\T\in \mathbb{C}^{K},
\end{align}with 
\begin{equation}
\label{eq:array_response_b}
    \vect{b}(\gamma_k) = \left[e^{-j2\gamma_k(N+1)},\ldots,e^{-j2\gamma_k(J-N)} \right]^\T.
\end{equation}
\begin{remark}
  To avoid DoA ambiguity in \eqref{eq:array_response_b}, the antenna spacing must satisfy $d\leq \lambda/4$. 
\end{remark}

Averaging over the outer products of the $J$ subvectors with themselves, we obtain 
\begin{equation}
    \bar{\vect{R}} =\frac{1}{J}\sum_{i=1}^J \bar{\vect{y}}_i\bar{\vect{y}}_i^\H= \frac{1}{J}  \vect{B}\underbrace{\left(\sum_{i=1}^J\vect{p}_i \vect{p}_i^\H \right)}_{\triangleq \vect{R}_p}\vect{B}^\H = \frac{1}{J} \vect{B} \vect{R}_p \vect{B}^\H, 
\end{equation} and the eigenvalue decomposition of $\bar{\vect{R}}$ yields 
\begin{equation}
   \bar{\vect{R}}  = \bar{\vect{U}}_s \bar{\bl{\Sigma}}_s \bar{\vect{U}}_s^\H + \bar{\vect{U}}_n \bar{\bl{\Sigma}}_n \bar{\vect{U}}_n^\H.  
\end{equation} Employing the MUSIC algorithm, the $K$ DoAs can be estimated by finding the $K$ peaks of the 1D spectrum
\begin{equation}
\label{eq:spectrum_angle}
   f(\theta) = \frac{1}{\vect{b}^\H(\gamma) \bar{\vect{U}}_n \bar{\vect{U}}_n^\H \vect{b}(\gamma) }. 
\end{equation}The above MUSIC-based algorithm was first developed in \citep{He2012efficient}. We refer to this algorithm as \textit{modified MUSIC} in the rest of the chapter. 
\begin{remark}
 The modified MUSIC algorithm described above only works when $\mathrm{rank}(\vect{B}) = K$ and $\mathrm{rank}(\vect{R}_p) = K$. The former is satisfied when $K < 2N+2-J$. Furthermore, matrix $\vect{R}_p$ is sum of $J$ rank-one matrices; thus, it becomes a full-rank matrix when $J> K$. These two conditions imply that $K < N+1$. Therefore, the number of resolvable sources is at most $K = N = \frac{M-1}{2}$ with the modified MUSIC algorithm.  
\end{remark}

After estimating the DoAs, the standard MUSIC algorithm is applied for finding the corresponding ranges. In particular, for each DoA estimate $\hat{\theta}_k$, the range $r_k$ is estimated as 
\begin{equation}
\label{eq:range_estimation}
    \hat{r}_k = \argmax{r}\,\frac{1}{\vect{a}^\H(\hat{\theta}_k,r)\vect{U}_n \vect{U}_n^\H \vect{a}(\hat{\theta}_k,r)}, \quad k=1,\ldots,K.
\end{equation}
\begin{figure}
	\centering
	\begin{subfigure}{\columnwidth}
		\centering
		\includegraphics[width=0.88\columnwidth]{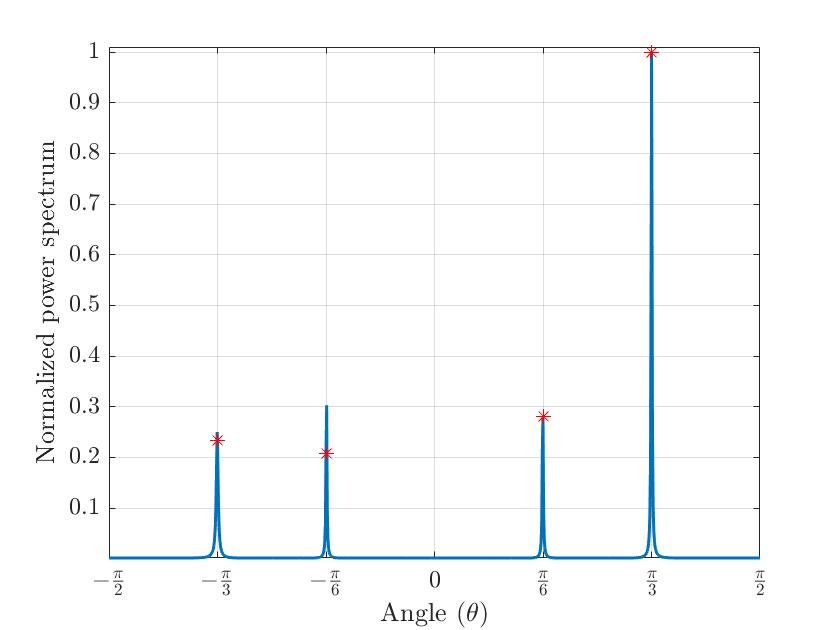}
		\caption{DoA estimation}
		\label{fig:Modified_MUSIC_DoA}
  \vspace{-1mm}
	\end{subfigure}
	\begin{subfigure}{\columnwidth}
		\centering
		\includegraphics[width=0.88\columnwidth]{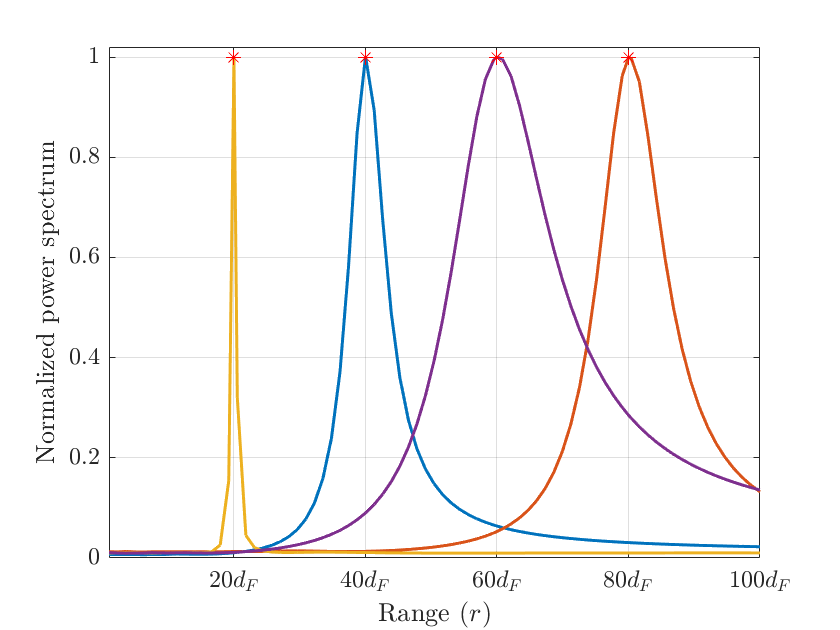}
		\caption{Range estimation}
		\label{fig:Modified_MUSIC_Range}
   \vspace{-3mm}
	\end{subfigure}
 \caption{The normalized power spectra when there are four sources located at $(-\pi/3,60\,d_\F)$, $(-\pi/6,80\,d_\F)$, $(\pi/6,20\,d_\F)$, and $(\pi/3,40\,d_\F)$. First, the DoAs are estimated as the four highest peaks of the spectrum in \eqref{eq:spectrum_angle} and then, the corresponding ranges are obtained via the standard MUSIC algorithm. True locations are marked by red stars. }
 \label{fig:Modified_MUSIC}
\end{figure}Figure~\ref{fig:Modified_MUSIC} shows the result of DoA and range estimation using the modified MUSIC algorithm. The number of antennas is assumed to be $M = 51$ with $d=\lambda/4$ inter-antenna spacing, and the number of overlapping subvectors is set as $J = 20$. $L = 100$ time samples are used for the estimation. There are $K = 4$ sources with their angle-range pairs given by $(-\pi/3,60\,d_\F)$, $(-\pi/6,80\,d_\F)$, $(\pi/6,20\,d_\F)$, and $(\pi/3,40\,d_\F)$. The SNR is $0\,$dB per antenna and per sample. 
Figure~\ref{fig:Modified_MUSIC}(a) illustrates  the power spectrum in \eqref{eq:spectrum_angle} whose peak value is normalized to 1. It is observed that the DoAs have been accurately identified. Based on the identified DoAs, the estimator in \eqref{eq:range_estimation} finds the corresponding ranges of the sources. The results, depicted in Figure~\ref{fig:Modified_MUSIC}(b), show the effectiveness of the MUSIC in estimating the range of near-field sources.   

The number of sources to be localized is limited to $K \leq N = \frac{M-1}{2}$ with the modified MUSIC algorithm, thus incurring an array aperture loss. Next, we describe another DoA estimation approach based on the generalized ESPRIT algorithm presented in Section~\ref{sec:gen_ESPRIT}, which can alleviate this issue. 

\subsubsection{Generalized ESPRIT for Near-Field Localization}

Building on the symmetric structure of the array response vector in \eqref{eq:array_response_symmetric}, the antenna array can be divided into two subarrays each having $J < M$ elements, where the first subarray contains the first $J$ antennas in ascending order and the second subarray consists of the last $J$ antennas in descending order. Based on \eqref{eq:received_signal_symmetric}, the signals received at the two subarrays are given by 
\begin{align}
   \vect{y}_1(t) &= \left[y_{-N}(t),y_{-N+1}(t),\ldots,y_{-N + (J-1)}(t)\right]^\T \nonumber\\
   &= \vect{A}_1 \diag\left(e^{j\psi_1},\ldots,e^{j\psi_K}\right)\vect{s}(t) + \vect{n}_1(t), \\  
   \vect{y}_2(t) & = \left[y_N(t),y_{N-1}(t),\ldots,y_{N-(J-1)}(t) \right]^\T \nonumber\\
   &= \vect{A}_2 \diag\left(e^{j\psi_1},\ldots,e^{j\psi_K}\right)\vect{s}(t) + \vect{n}_2(t),
\end{align}
where 
\begin{align}
   \vect{A}_1 = \left[\vect{a}_1(\theta_1,r_1),\ldots,\vect{a}_1(\theta_K,r_K)\right]\in \mathbb{C}^{J\times K}, \\
   \vect{A}_2 = \left[\vect{a}_2(\theta_1,r_1),\ldots,\vect{a}_2(\theta_K,r_K)\right]\in \mathbb{C}^{J\times K},
\end{align}and the array response vectors of the subarrays are given by 
\begin{align}
    \vect{a}_1(\theta_k,r_k) &= \left[e^{j\left(-N\gamma_k + N^2 \phi_k\right) },\ldots,e^{j\left((-N+J-1)\gamma_k + (-N+J-1)^2 \phi_k\right) } \right]^\T, \\
    \vect{a}_2(\theta_k,r_k) & = \left[e^{j\left(N\gamma_k + N^2 \phi_k\right)},\ldots, e^{j\left((N - J+1)\gamma_k + (N-J+1)^2 \phi_k\right)} \right]^\T.
\end{align}The array response matrix $\vect{A}$ can thus be divided as 
\begin{equation}
\label{eq:array_response_split}
    \vect{A}=
\begin{bmatrix}
\vect{A}_1 \\
 \mathrm{Last~} (M-J) \mathrm{~rows}
\end{bmatrix} = \begin{bmatrix}
\mathrm{First~} (M-J) \mathrm{~rows} \\
 \vect{E}\vect{A}_2,
\end{bmatrix} 
\end{equation}where $\vect{E}$ is the exchange matrix having ones on its anti-diagonal and zeros elsewhere, and $\vect{E}^2 = \vect{I}_J$. The array response vector of the second subarray can be represented as 
\begin{equation}
   \vect{A}_2 = \left[\vect{D}(\theta_1)\vect{a}_1(\theta_1,r_1),\ldots,\vect{D}(\theta_K)\vect{a}_1(\theta_K,r_K) \right], 
\end{equation}where 
\begin{equation}
    \vect{D}(\theta_k)=
\begin{bmatrix}
e^{j \frac{4\pi d}{\lambda}\sin (\theta_k) N} &  & 0 \\
 & \ddots &  \\
0 &  & e^{j \frac{4\pi d}{\lambda}\sin (\theta_k) (N-J+1)}
\end{bmatrix}.
\end{equation} Recall from \eqref{eq:Us-AC} that $\vect{U}_s = \vect{AC}$. Similar to \eqref{eq:array_response_split}, we can split $\vect{U}_s$ and $\vect{AC}$ as 
\begin{align}
\vect{U}_s &=
\begin{bmatrix}
\vect{U}_{s1} \\
 \mathrm{Last~} (M-J) \mathrm{~rows}
\end{bmatrix} = \begin{bmatrix}
\mathrm{First~} (M-J) \mathrm{~rows} \\
 \vect{U}_{s2}
\end{bmatrix}, \\
\vect{AC} &=
\begin{bmatrix}
\vect{A}_1 \vect{C} \\
 \mathrm{Last~} (M-J) \mathrm{~rows}
\end{bmatrix} = \begin{bmatrix}
\mathrm{First~} (M-J) \mathrm{~rows} \\
 \vect{E}\vect{A}_2 \vect{C}
\end{bmatrix},
\end{align}which implies that 
\begin{align}
   \vect{U}_{s1} &= \vect{A}_1 \vect{C}, \\ 
   \vect{E}\vect{U}_{s2} &= \vect{A}_2 \vect{C}.
\end{align}Following the generalized ESPRIT method in Section~\ref{sec:gen_ESPRIT}, the diagonal matrix $\bl{\Psi}$ is introduced as
\begin{equation}
    \bl{\Psi}(\theta)=
\begin{bmatrix}
e^{j\frac{4\pi d}{\lambda}\sin (\theta) N} &  & 0 \\
 & \ddots &  \\
0 &  & e^{j \frac{4\pi d}{\lambda}\sin (\theta) (N-J+1)}
\end{bmatrix}
\end{equation} and matrix $\vect{F}(\theta)$ is formed as 
\begin{equation}
    \vect{F} (\theta) = \vect{E}\vect{U}_{s2} - \bl{\Psi}(\theta) \vect{U}_{s1} = \vect{Q}\vect{C},
\end{equation}with 
\begin{align}
    \vect{Q} = \left[\left(\vect{D}(\theta_1) - \bl{\Psi}(\theta)\right)\vect{a}_1(\theta_1,r_1),\ldots, \left(\vect{D}(\theta_K) - \bl{\Psi}(\theta)\right)\vect{a}_1(\theta_K,r_K)\right].
\end{align} 
The matrix $\vect{F}(\theta_k)$ is rank-deficient, which means that for an arbitrary full column rank matrix $\vect{W} \in \mathbb{C}^{J\times K}$ the matrix $\vect{W}^\H \vect{F}(\theta_k)$ is singular and its determinant is zero. 
 The DoAs of the sources can therefore be found as the $K$ highest peaks of the spectrum 
\begin{equation}
\label{eq:spectrum_gen_ESPRIT_nearfield}
f(\theta) = \frac{1}{\det \left(\vect{W}^\H  \vect{F}(\theta) \right)}.
\end{equation}
\begin{figure}
	\centering
	\begin{subfigure}{\columnwidth}
		\centering
		\includegraphics[width=0.88\columnwidth]{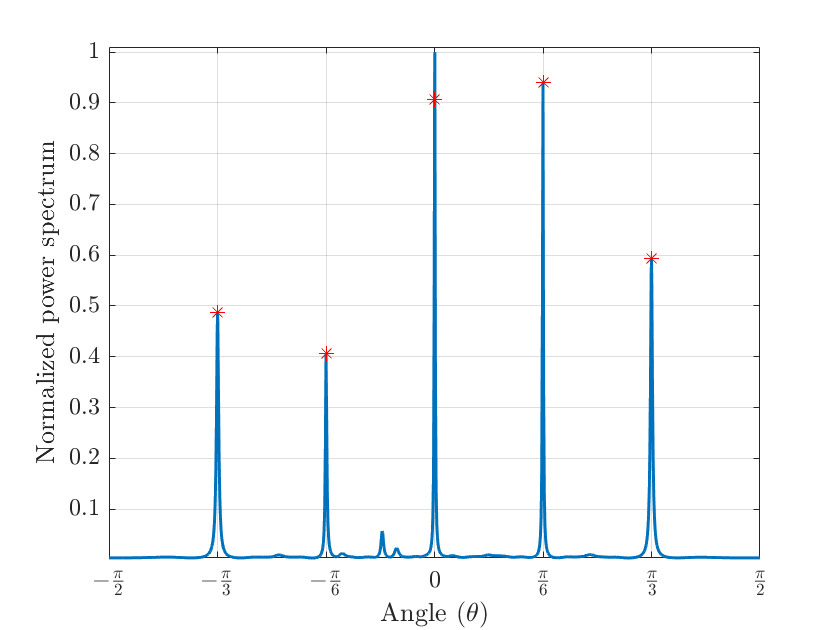}
		\caption{DoA estimation}
		\label{fig:Gen_ESP_DoA}
  \vspace{-1mm}
	\end{subfigure}
	\begin{subfigure}{\columnwidth}
		\centering
		\includegraphics[width=0.88\columnwidth]{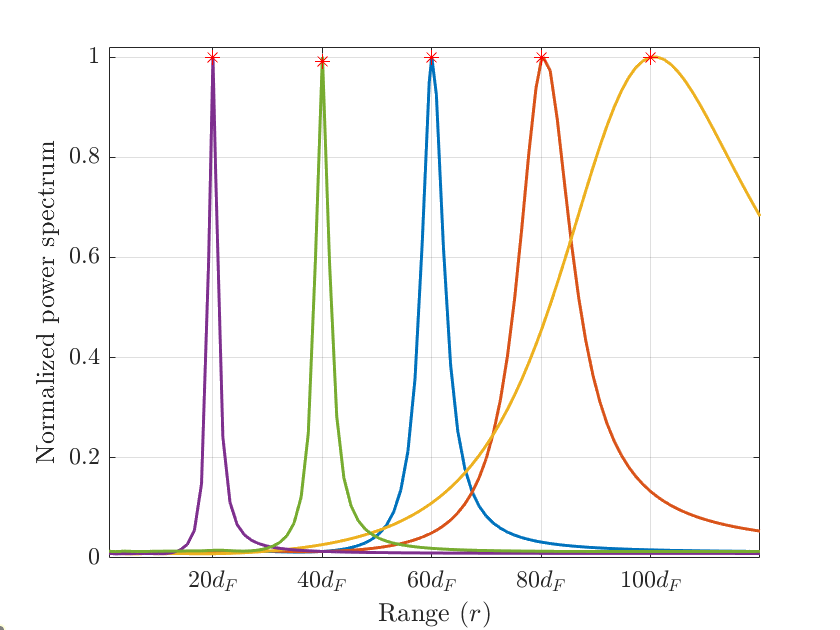}
		\caption{Range estimation}
		\label{fig:Gen_ESP_Range}
  \vspace{-3mm}
	\end{subfigure}
 \caption{The normalized power spectra when there are five sources located at $(-\pi/3,20\,d_\F)$, $(-\pi/6,40\,d_\F)$, $(0,60\,d_\F)$, $(\pi/6,80\,d_\F)$, and $(\pi/3,100\,d_\F)$. First, the DoAs are estimated as the five highest peaks of the spectrum in \eqref{eq:spectrum_gen_ESPRIT_nearfield} and then, the corresponding ranges are obtained via the standard MUSIC algorithm. True locations are marked by red stars.}
 \label{fig:Gen_ESP}
\end{figure}This generalized ESPRIT-based approach for finding the DoAs of near-field sources was first developed in \citep{Zhi2007nearfield}.
\begin{remark}
Using the generalized ESPRIT method, $K \leq J$ sources can be resolved, and because $J < M$, the maximum number of resolvable sources with this method is $K = M-1$, which is twice the number of resolvable sources compared with the modified MUSIC algorithm. 
\end{remark}

After estimating the DoAs of the sources as $\hat{\theta}_1,\ldots,\hat{\theta}_K$, the ranges can be found from \eqref{eq:range_estimation}.

Figure~\ref{fig:Gen_ESP} shows the normalized spectra for DoA and range, where generalized ESPRIT is utilized for DoA estimation and standard MUSIC is used for range estimation. The number of antennas is $M=51$ with $d = \lambda/4$ spacing and the number of samples is set to be $L = 100$. Furthermore, the antenna array is divided into two subarrays, each having $J = 50$ antennas.  There are $K = 5$ sources with the angle-range pairs of $(-\pi/3,20\,d_\F)$, $(-\pi/6,40\,d_\F)$, $(0,60\,d_\F)$, $(\pi/6,80\,d_\F)$, and $(\pi/3,100\,d_\F)$. We assume that the number of sources, $K$, has already been specified \citep{Wax1985}. The SNR is set as $0\,$dB per antenna and per sample. We set $\vect{W} = \vect{F}(\theta)$ since $\vect{F}(\theta)$ is full column rank when $\theta \neq \theta_k,~k = 1,\ldots,K$.
From Figure~\ref{fig:Gen_ESP}(a), it can be observed that the generalized ESPRIT algorithm effectively resolves the DoAs of the five sources. While there are also some peaks at other angles, their power is much smaller than the power of the peaks at the true DoAs. Since the DoAs are accurately estimated in the first step, the MUSIC algorithm can find the ranges of near-field sources with a good accuracy, as observed in Figure~\ref{fig:Gen_ESP}(b). 

\begin{figure}
    \centering
    \includegraphics[width = 0.88\columnwidth]{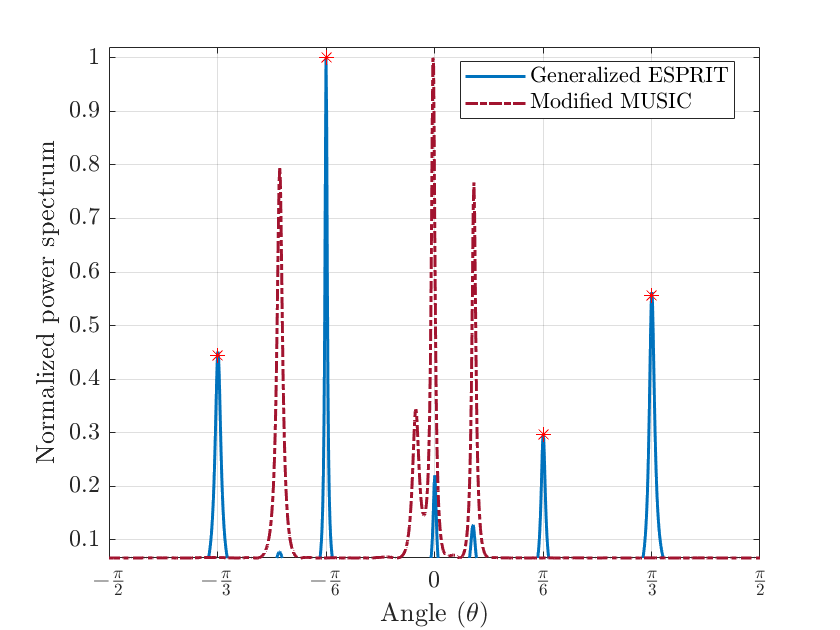}
    \caption{The normalized power spectra for the modified MUSIC and generalized ESPRIT algorithms when the correlation coefficient between source signals is $0.9$. The setup is the same as in Figure~\ref{fig:Modified_MUSIC}. True angles are marked by red stars.}
    \label{fig:correlated_sources}
\end{figure}

We have so far assumed that the signals arriving from different sources at the ULA are perfectly uncorrelated. However, in practice, the existence of multipath propagation due to reflection, refraction, and scattering may result in coherent signals that are strongly correlated. 
We now consider the correlation between source signals and compare the performance of the modified MUSIC and generalized ESPRIT algorithms in the presence of correlation. We consider the same scenario as the one in Figure~\ref{fig:Modified_MUSIC} and the correlation coefficient between each pair of sources is assumed to be $0.9$. Figure~\ref{fig:correlated_sources} shows the normalized power spectra for the modified MUSIC and generalized ESPRIT algorithms. We can observe that the modified MUSIC algorithm fails to estimate the DoAs of the sources and produces four peaks that are far from the true DoAs. On the other hand, the peaks generated by the generalized ESPRIT algorithm are close to the true DoAs, which suggests that this algorithm can better handle source correlations compared to the modified MUSIC algorithm. 

\section{Conclusions}
\label{sec:conclusion}
As Massive MIMO systems expand to accommodate more users through spatial multiplexing, the increased size of antenna arrays is also beneficial for localization services. This is because having more antennas enhances spatial resolution and allows for more accurate localization of closely spaced sources. In this chapter, we covered two important subspace-based localization algorithms, namely MUSIC and ESPRIT, for localizing far-field sources, and introduced their variants that are employed for near-field localization. 
These algorithms can be directly implemented on Massive MIMO base stations, and the source signals might be uplink data transmitted by user devices.
We observed that both MUSIC and ESPRIT provide high-resolution DoA estimation, with MUSIC slightly outperforming ESPRIT at the expense of increased complexity, which stems from evaluating the spatial spectrum at various grid points corresponding to possible DoAs. Unlike ESPRIT, the MUSIC algorithm can be directly applied to near-field localization where the conventional 1D grid of possible DoAs is extended to a 2D grid involving both DoA and range values and a 2D grid search is subsequently performed for localizing the sources. We further introduced two methods based on MUSIC and ESPRIT algorithms that, by decoupling DoA and range estimation problems, reduce the complexity of the 2D MUSIC algorithm. 
The presented localization algorithms find applications in various fields where determining the location of signal sources is crucial. These fields include radar systems, wireless communications, sonar systems, astronomy, seismic monitoring, etc.

As a promising future research direction, one could explore the application of the presented algorithms to the scenario where the antenna array is a uniform planar array and the near-field localization problem involves the estimation of azimuth and elevation DoAs as well as the range of the sources.  In such cases, a new search dimension is added to the localization problem, which further increases the computational complexity and necessitates the development of efficient algorithms.  Inspired by the symmetry-based near-field localization algorithms, it is interesting to investigate new ways for decomposing the localization problem into separate subproblems of azimuth, elevation, and range estimation problems, without compromising the estimation accuracy. Furthermore, as the same signals can be used for both localization and communication purposes, the newly emerged topic of joint localization and communication is worth further investigation.  

In addition to the methods discussed in this chapter, other widely employed localization techniques are available in the literature. Table \ref{tab:other_techniques} presents some of these techniques along with relevant references.

\begin{table}
\renewcommand{\arraystretch}{1.3}
\captionsetup{format =hang}%
\caption{\mbox{Other widely-used localization techniques.}\label{tab:other_techniques}}{
\LARGE
\centering
\resizebox{\columnwidth}{!}{%
\begin{tabular}{|l|l|}
\hline
\rowcolor{lightgray}\multicolumn{1}{|c|}{\textbf{Technique}} & \multicolumn{1}{c|}{\textbf{References}}\\
\hline
Time difference of arrival  &  \citep{Wei2010Multi},\citep{Wang2017TDOA},\citep{Sun2019Solution}\\
\hline
Linear prediction & \citep{Grosicki},\citep{Zuo2019}\\
\hline
Maximum likelihood & \citep{Chen2002},\citep{Cheng2022an}\\
\hline
Sparse signal reconstruction & \citep{Malioutov},\citep{Wang2021Mixed}\\
\hline
\end{tabular}}}{}
\end{table}

\backmatter

\bibliographystyle{plainnat}
\bibliography{wiley}%

%% file: main.bbl
\begin{thebibliography}{32}
\providecommand{\natexlab}[1]{#1}
\providecommand{\url}[1]{\texttt{#1}}
\expandafter\ifx\csname urlstyle\endcsname\relax
  \providecommand{\doi}[1]{doi: #1}\else
  \providecommand{\doi}{doi: \begingroup \urlstyle{rm}\Url}\fi

\bibitem[Bj{\"o}rnson et~al.(2017)Bj{\"o}rnson, Hoydis, Sanguinetti, et~al.]{bjornson2017massive}
Emil Bj{\"o}rnson, Jakob Hoydis, Luca Sanguinetti, et~al.
\newblock Massive {MIMO} networks: Spectral, energy, and hardware efficiency.
\newblock \emph{Foundations and Trends{\textregistered} in Signal Processing}, 11\penalty0 (3-4):\penalty0 154--655, 2017.

\bibitem[Chen et~al.(2023)Chen, Keskin, Sakhnini, Decarli, Pollin, Dardari, and Wymeersch]{chen20236g}
Hui Chen, Musa~Furkan Keskin, Adham Sakhnini, Nicol{\'o} Decarli, Sofie Pollin, Davide Dardari, and Henk Wymeersch.
\newblock {6G} localization and sensing in the near field: {Fundamentals}, opportunities, and challenges.
\newblock \emph{arXiv preprint arXiv:2308.15799}, 2023.

\bibitem[Chen et~al.(2002)Chen, Hudson, and Yao]{Chen2002}
Joe~C Chen, Ralph~E Hudson, and Kung Yao.
\newblock Maximum-likelihood source localization and unknown sensor location estimation for wideband signals in the near-field.
\newblock \emph{IEEE Transactions on Signal Processing}, 50\penalty0 (8):\penalty0 1843--1854, 2002.
\newblock \doi{10.1109/TSP.2002.800420}.

\bibitem[Cheng et~al.(2022)Cheng, Liu, Wu, and Zhang]{Cheng2022an}
Cheng Cheng, Songyong Liu, Hongzhuang Wu, and Ying Zhang.
\newblock An efficient maximum-likelihood-like algorithm for near-field coherent source localization.
\newblock \emph{IEEE Transactions on Antennas and Propagation}, 70\penalty0 (7):\penalty0 6111--6116, 2022.
\newblock \doi{10.1109/TAP.2022.3161269}.

\bibitem[Friedlander(2019)]{8736783}
Benjamin Friedlander.
\newblock Localization of signals in the near-field of an antenna array.
\newblock \emph{IEEE Transactions on Signal Processing}, 67\penalty0 (15):\penalty0 3885--3893, 2019.
\newblock \doi{10.1109/TSP.2019.2923164}.

\bibitem[Gao and Gershman(2005)]{Gao2005generalized}
Feifei Gao and Alex~B Gershman.
\newblock A generalized {ESPRIT} approach to direction-of-arrival estimation.
\newblock \emph{IEEE Signal Processing Letters}, 12\penalty0 (3):\penalty0 254--257, 2005.
\newblock \doi{10.1109/LSP.2004.842276}.

\bibitem[Grosicki et~al.(2005)Grosicki, Abed-Meraim, and Hua]{Grosicki}
Emmanu{\`e}le Grosicki, Karim Abed-Meraim, and Yingbo Hua.
\newblock A weighted linear prediction method for near-field source localization.
\newblock \emph{IEEE Transactions on Signal Processing}, 53\penalty0 (10):\penalty0 3651--3660, 2005.
\newblock \doi{10.1109/TSP.2005.855100}.

\bibitem[He et~al.(2012)He, Swamy, and Ahmad]{He2012efficient}
Jin He, MNS Swamy, and M~Omair Ahmad.
\newblock Efficient application of {MUSIC} algorithm under the coexistence of far-field and near-field sources.
\newblock \emph{IEEE Transactions on Signal Processing}, 60\penalty0 (4):\penalty0 2066--2070, 2012.
\newblock \doi{10.1109/TSP.2011.2180902}.

\bibitem[Krim and Viberg(1996)]{Krim1996a}
Hamid Krim and Mats Viberg.
\newblock Two decades of array signal processing research: the parametric approach.
\newblock 13\penalty0 (4):\penalty0 67--94, 1996.

\bibitem[Liang and Liu(2009)]{liang2009passive}
Junli Liang and Ding Liu.
\newblock Passive localization of mixed near-field and far-field sources using two-stage {MUSIC} algorithm.
\newblock \emph{IEEE Transactions on Signal Processing}, 58\penalty0 (1):\penalty0 108--120, 2009.

\bibitem[Liu and Sun(2013)]{Liu2013Efficient}
Guohong Liu and Xiaoying Sun.
\newblock Efficient method of passive localization for mixed far-field and near-field sources.
\newblock \emph{IEEE Antennas and Wireless Propagation Letters}, 12:\penalty0 902--905, 2013.
\newblock \doi{10.1109/LAWP.2013.2273451}.

\bibitem[Liu et~al.(2023)Liu, Haardt, Greco, Mecklenbräuker, and Willett]{10144727}
Wei Liu, Martin Haardt, Maria~S. Greco, Christoph~F. Mecklenbräuker, and Peter Willett.
\newblock Twenty-five years of sensor array and multichannel signal processing: {A} review of progress to date and potential research directions.
\newblock \emph{IEEE Signal Processing Magazine}, 40\penalty0 (4):\penalty0 80--91, 2023.
\newblock \doi{10.1109/MSP.2023.3258060}.

\bibitem[Malioutov et~al.(2005)Malioutov, Cetin, and Willsky]{Malioutov}
Dmitry Malioutov, M{\"u}jdat Cetin, and Alan~S Willsky.
\newblock A sparse signal reconstruction perspective for source localization with sensor arrays.
\newblock \emph{IEEE Transactions on Signal Processing}, 53\penalty0 (8):\penalty0 3010--3022, 2005.
\newblock \doi{10.1109/TSP.2005.850882}.

\bibitem[Pesavento et~al.(2023)Pesavento, Trinh-Hoang, and Viberg]{pesavento2023three}
Marius Pesavento, Minh Trinh-Hoang, and Mats Viberg.
\newblock Three more decades in array signal processing research: {An} optimization and structure exploitation perspective.
\newblock \emph{IEEE Signal Processing Magazine}, 40\penalty0 (4):\penalty0 92--106, 2023.

\bibitem[Ramezani and Bj{\"o}rnson(2023)]{ramezani2022bookchapter}
Parisa Ramezani and Emil Bj{\"o}rnson.
\newblock Near-field beamforming and multiplexing using extremely large aperture arrays.
\newblock In \emph{Fundamentals of 6G Communications and Networking}, pages 317--349. Springer, 2023.

\bibitem[Ramezani et~al.(2023)Ramezani, Kosasih, Irshad, and Björnson]{ramezani2023magazine}
Parisa Ramezani, Alva Kosasih, Amna Irshad, and Emil Björnson.
\newblock Exploiting the depth and angular domains for massive near-field spatial multiplexing.
\newblock \emph{IEEE BITS the Information Theory Magazine}, pages 1--12, 2023.
\newblock \doi{10.1109/MBITS.2023.3322670}.

\bibitem[Rockah and Schultheiss(1987)]{Rockah1987}
Yosef Rockah and Peter Schultheiss.
\newblock Array shape calibration using sources in unknown locations--part {II}: Near-field sources and estimator implementation.
\newblock \emph{IEEE Transactions on Acoustics, Speech, and Signal Processing}, 35\penalty0 (6):\penalty0 724--735, 1987.
\newblock \doi{10.1109/TASSP.1987.1165222}.

\bibitem[Roy et~al.(1986)Roy, Paulraj, and Kailath]{Roy1986}
Robert Roy, Arogyaswami Paulraj, and Thomas Kailath.
\newblock {ESPRIT}--a subspace rotation approach to estimation of parameters of cisoids in noise.
\newblock \emph{IEEE Transactions on Acoustics, Speech, and Signal Processing}, 34\penalty0 (5):\penalty0 1340--1342, 1986.
\newblock \doi{10.1109/TASSP.1986.1164935}.

\bibitem[Schmidt(1986)]{Schmidt1986}
Ralph Schmidt.
\newblock Multiple emitter location and signal parameter estimation.
\newblock \emph{IEEE Transactions on Antennas and Propagation}, 34\penalty0 (3):\penalty0 276--280, 1986.
\newblock \doi{10.1109/TAP.1986.1143830}.

\bibitem[Stoica et~al.(2005)Stoica, Moses, et~al.]{stoica2005spectral}
Petre Stoica, Randolph~L Moses, et~al.
\newblock \emph{Spectral analysis of signals}, volume 452.
\newblock Pearson Prentice Hall Upper Saddle River, NJ, 2005.

\bibitem[Sun et~al.(2019)Sun, Ho, and Wan]{Sun2019Solution}
Yimao Sun, KC~Ho, and Qun Wan.
\newblock Solution and analysis of {TDOA} localization of a near or distant source in closed form.
\newblock \emph{IEEE Transactions on Signal Processing}, 67\penalty0 (2):\penalty0 320--335, 2019.
\newblock \doi{10.1109/TSP.2018.2879622}.

\bibitem[Wang et~al.(2012)Wang, Liu, and Sun]{Wang2021Mixed}
Bo~Wang, Juanjuan Liu, and Xiaoying Sun.
\newblock Mixed sources localization based on sparse signal reconstruction.
\newblock \emph{IEEE Signal Processing Letters}, 19\penalty0 (8):\penalty0 487--490, 2012.
\newblock \doi{10.1109/LSP.2012.2204248}.

\bibitem[Wang et~al.(2013)Wang, Zhao, and Liu]{6451125}
Bo~Wang, Yanping Zhao, and Juanjuan Liu.
\newblock Mixed-order {MUSIC} algorithm for localization of far-field and near-field sources.
\newblock \emph{IEEE Signal Processing Letters}, 20\penalty0 (4):\penalty0 311--314, 2013.
\newblock \doi{10.1109/LSP.2013.2245503}.

\bibitem[Wang and Ho(2017)]{Wang2017TDOA}
Yue Wang and K.~C. Ho.
\newblock {TDOA} positioning irrespective of source range.
\newblock \emph{IEEE Transactions on Signal Processing}, 65\penalty0 (6):\penalty0 1447--1460, 2017.
\newblock \doi{10.1109/TSP.2016.2630030}.

\bibitem[Wax and Kailath(1983)]{Wax1983}
Mati Wax and Thomas Kailath.
\newblock Optimum localization of multiple sources by passive arrays.
\newblock \emph{IEEE Transactions on Acoustics, Speech, and Signal Processing}, 31\penalty0 (5):\penalty0 1210--1217, 1983.
\newblock \doi{10.1109/TASSP.1983.1164183}.

\bibitem[Wax and Kailath(1985)]{Wax1985}
Mati Wax and Thomas Kailath.
\newblock Detection of signals by information theoretic criteria.
\newblock \emph{IEEE Transactions on Acoustics, Speech, and Signal Processing}, 33\penalty0 (2):\penalty0 387--392, 1985.
\newblock \doi{10.1109/TASSP.1985.1164557}.

\bibitem[Wei et~al.(2010)Wei, Peng, Wan, Chen, and Ye]{Wei2010Multi}
He-Wen Wei, Rong Peng, Qun Wan, Zhang-Xin Chen, and Shang-Fu Ye.
\newblock Multidimensional scaling analysis for passive moving target localization with {TDOA} and {FDOA} measurements.
\newblock \emph{IEEE Transactions on Signal Processing}, 58\penalty0 (3):\penalty0 1677--1688, 2010.
\newblock \doi{10.1109/TSP.2009.2037666}.

\bibitem[Zhang et~al.(2023)Zhang, Li, Qi, and Li]{9960789}
Jiawen Zhang, Yingsong Li, Junwei Qi, and Yibing Li.
\newblock Symmetric extended nested array for passive localization of mixture near- and far-field sources.
\newblock \emph{IEEE Transactions on Circuits and Systems II: Express Briefs}, 70\penalty0 (3):\penalty0 1244--1248, 2023.
\newblock \doi{10.1109/TCSII.2022.3223940}.

\bibitem[Zheng et~al.(2021)Zheng, Fu, Wang, and So]{9132699}
Zhi Zheng, Mingcheng Fu, Wen-Qin Wang, and Hing~Cheung So.
\newblock Symmetric displaced coprime array configurations for mixed near- and far-field source localization.
\newblock \emph{IEEE Transactions on Antennas and Propagation}, 69\penalty0 (1):\penalty0 465--477, 2021.
\newblock \doi{10.1109/TAP.2020.3005203}.

\bibitem[Zhi and Chia(2007)]{Zhi2007nearfield}
Wanjun Zhi and Michael Yan-Wah Chia.
\newblock Near-field source localization via symmetric subarrays.
\newblock \emph{IEEE Signal Processing Letters}, 14\penalty0 (6):\penalty0 409--412, 2007.
\newblock \doi{10.1109/LSP.2006.888390}.

\bibitem[Ziskind and Wax(1988)]{Ziskind1988}
Ilan Ziskind and Mati Wax.
\newblock Maximum likelihood localization of multiple sources by alternating projection.
\newblock \emph{IEEE Transactions on Acoustics, Speech, and Signal Processing}, 36\penalty0 (10):\penalty0 1553--1560, 1988.
\newblock \doi{10.1109/29.7543}.

\bibitem[Zuo et~al.(2019)Zuo, Xin, Liu, Zheng, Ohmori, and Sano]{Zuo2019}
Weiliang Zuo, Jingmin Xin, Wenyi Liu, Nanning Zheng, Hiromitsu Ohmori, and Akira Sano.
\newblock Localization of near-field sources based on linear prediction and oblique projection operator.
\newblock \emph{IEEE Transactions on Signal Processing}, 67\penalty0 (2):\penalty0 415--430, 2019.
\newblock \doi{10.1109/TSP.2018.2883034}.

\end{thebibliography}
